\documentclass[final,authoryear,3p,times]{elsarticle}
\usepackage{afterpage}
\usepackage{algorithm}
\usepackage{algpseudocode}
\usepackage{amsmath}
\usepackage{amssymb}
\usepackage{amsthm}
\usepackage{bm}
\usepackage{booktabs}
\usepackage{calc}
\usepackage{caption}
\usepackage{subcaption}
\usepackage[table]{xcolor} % Required for \rowcolor
\usepackage{amsmath,amssymb}
\usepackage{booktabs}
\usepackage{tabularx}

\usepackage{enumerate}
\usepackage{graphicx}
\usepackage{ifthen}
\usepackage[parfill]{parskip} % no indent after line break
\usepackage{multicol}
\usepackage{siunitx}
\usepackage{ulem}

\usepackage[colorlinks]{hyperref}
\usepackage[nameinlink,capitalise]{cleveref}
\crefname{appendix}{}{}
\usepackage{tikz}
\usetikzlibrary{arrows}
\usetikzlibrary{automata}
\usetikzlibrary{calc}
\usetikzlibrary{chains}
\usetikzlibrary{positioning}
\usetikzlibrary{shapes}

\usepackage{varwidth}

\usepackage{tcolorbox}
\newtcolorbox[auto counter]{problem}[2][]{colframe=blue!30, colback=blue!5, coltitle=black, title=Problem~\thetcbcounter~ #2,#1}

\usepackage{spverbatim} % like verbatim with linebreaks
\usepackage{listings} % like verbatim with linebreaks

\definecolor{Demiray}{rgb}{0.1725, 0.6275, 0.1725}
\definecolor{Gent}{rgb}{1.0, 0.4902, 0.0431}
\definecolor{Holzapfel}{rgb}{0.1216, 0.4667, 0.7059}
\definecolor{MooneyRivlin}{rgb}{0.8392, 0.1529, 0.1569}
\definecolor{Ogden}{rgb}{0.5804, 0.4039, 0.7412}
\definecolor{BlatzKo}{rgb}{0.5490, 0.3373, 0.2941}
\definecolor{NeoHooke}{rgb}{0.7373, 0.7412, 0.1333}

\usepackage{cleveref}

\edef\svtheparindent{\the\parindent}
\usepackage{parskip}
\usepackage{ulem}
\parindent=\svtheparindent\relax

\newcommand{\MF}{Material Fingerprinting}

\definecolor{ForestGreen}{RGB}{34,139,34}
\definecolor{InternationalOrange}{rgb}{1.0, 0.31, 0.0}
\definecolor{WineRed}{RGB}{139,0,0}

\newcommand{\CHANGE}[1]{#1}

\newcommand{\boldface}[1]{\boldsymbol{#1}}  % italic (slanted)
\newcommand{\bfa}{\boldface{a}}

\newcommand{\bfe}{\boldface{e}}
\newcommand{\bff}{\boldface{f}}

\newcommand{\bfr}{\boldface{r}}

\newcommand{\bfC}{\boldface{C}}

\newcommand{\bfF}{\boldface{F}}

\newcommand{\bfH}{\boldface{H}}
\newcommand{\bfI}{\boldface{I}}

\newcommand{\bfM}{\boldface{M}}

\newcommand{\bfP}{\boldface{P}}
\newcommand{\bfQ}{\boldface{Q}}

\newcommand{\bfY}{\boldface{Y}}

\newcommand{\bfalpha}{\boldsymbol{\alpha}}

\newcommand{\bftheta}{\boldsymbol{\theta}}

\newcommand{\bfLambda}{\boldsymbol{\Lambda}}

\newcommand{\bfPhi}{\boldsymbol{\Phi}}

\newcommand{\bfOmega}{\boldsymbol{\Omega}}

\newcommand{\Rset}{\mathbb{R}}

\newcommand{\argmax}{\operatornamewithlimits{arg\ max}}

\newcommand{\be}{\begin{equation}}
\newcommand{\ee}{\end{equation}}
\newcommand{\bea}{\begin{equation}\begin{aligned}}
\newcommand{\eea}{\end{aligned}\end{equation}}
\newcommand{\beq}{\begin{eqnarray}}
\newcommand{\eeq}{\end{eqnarray}}
\newcommand{\bem}{\begin{multline}}
\newcommand{\eem}{\end{multline}}
\newcommand{\ba}{\begin{align}}
\newcommand{\ea}{\end{align}}
\newcommand{\bcase}{\left\{ \begin{array}{ll}}
\newcommand{\ecase}{\end{array} \right.}
\begin{document}

\begin{frontmatter}

\title{Adaptive Material Fingerprinting for the fast discovery of polyconvex feature combinations in isotropic and anisotropic hyperelasticity}
% \title{Adaptive Material Fingerprinting for the fast discovery of feature combinations in material modeling}
% \title{Adaptive Material Fingerprinting: \\ Fast discovery of dominant combinations of strain energy density features}
% \title{Adaptive Material Fingerprinting: \\ Fast material model discovery with tunable combinations of modeling features}

\fntext[contrib]{Moritz Flaschel and Hagen Holthusen contributed equally to this work.}
\cortext[cor1]{Correspondence: moritz.flaschel@fau.de}
\author[fau]{Moritz Flaschel\corref{cor1}\fnref{contrib}}
\author[fau]{Hagen Holthusen\fnref{contrib}}
\author[fau]{Denisa Martonová}
\author[fau,stan]{Ellen Kuhl}
\address[fau]{Institute of Applied Mechanics, Egerlandstraße 5, Friedrich-Alexander-Universität Erlangen-Nürnberg, 91058 Erlangen, Germany}
\address[stan]{Department of Mechanical Engineering, Stanford University, 440 Escondido Mall, California 94305, United States.}

\begin{abstract}

We recently proposed a method called \MF{} for the rapid discovery of mechanical material models that avoids solving continuous optimization problems.
\MF{} assumes that each material exhibits a unique response when subjected to a standardized experimental setup, which is interpreted as the material's mechanical fingerprint.
If a database of fingerprints is generated in an offline phase, a model for an unseen experimental measurement can be discovered in real time by comparing the experimentally measured fingerprint to the fingerprints in the database.
In our original contributions, the database comprised a fixed number of material models, each with a fixed number of parameters.
To increase the fitting flexibility of \MF{}, we propose an adaptive model database coupled with an iterative pattern recognition algorithm that refines the material model in each step.
This strategy enables \MF{} to discover arbitrary linear combinations of material models from the database, rather than being restricted to selecting a single model from a predefined set.
% and the maximum number of material parameters in the discovered material models is not fixed a priori.
In comparison to previous works on \MF{}, this enables the discovery of more complex models, such as multi-term Ogden models or the anisotropic Holzapfel-Gasser-Ogden model.
To design the adaptive database, we leverage sums of strain energy density feature functions that depend on isotropic and anisotropic invariants.
All modeling features satisfy fundamental physical constraints, and polyconvexity can be optionally enforced via a simple user-controlled switch.
% Specifically, we consider isotropic invariants $\lambda_i$ ($i=1,\dots,3$) and $\lambda_i\lambda_j$ ($i \neq j$), which describe principal stretches and principal area changes, respectively.
% Additionally, we consider anisotropic invariants $\lambda_{\bfa}$ and $A_{\bfa}$, which describe stretch in fiber direction and area change perpendicular to the fiber direction, respectively.
% These allow for a flexible description of the material behavior dependent on the isotropic invariants $\lambda_i$ ($i=1,\dots,3$) and $\lambda_i\lambda_j$ ($i \neq j$) and the anisotropic invariants $\lambda_{\bfa}$ and $A_{\bfa}$.
% This approach allows the determination of a minimal set of invariants required to fit a given experimental setup.
We test the method on experimental data stemming from mechanical tests of isotropic rubber materials and anisotropic animal skin tissue.
Our results show that the adaptive approach increases the fitting accuracy of \MF{} without significantly sacrificing the computational speed of the originally proposed method.
We found that the fitting accuracy of adaptive \MF{} is comparable to that of constitutive artificial neural networks while not relying on a time-consuming training process.
The hyperparameters of the adaptive method can be tuned to obtain sparse material models that are described by interpretable mathematical expressions.
\end{abstract}


\end{frontmatter}

%%%%%%%%%%%%%%%%%%%%%%%%%%%%%%%%%%%%%%%%%%%%%%%%%%%%%%%%%%%%%%%%%%%%%%%%%%%%%%%%%%%%%%%%%%%%%%%%%%%%%%%%%%%%%%%%%%%%%%%

% % < to be removed later
% \newpage
% To-Do:
% \begin{itemize}
    % \item alte todo's: adaptive MF mit ins pip package aufnehmen
    % \item alte todo's: mit NNLS und LASSO vergleichen
    % \item alte todo's: compressible case im Anhang skizzieren
% \end{itemize}

% \NOTE{Note: Ich benutze immer cref\{\} anstelle von ref\{\} oder eqref\{\}. cref\{\} erkennt automatisch ob etwas eine Gleichung, Figure oder Table ist.}

% \NOTE{Note: Let's use italic letters for natural number indices, e.g., $\lambda_i$ with $i=1,2,3$ and non-italic letters otherwise, e.g., $\bff^{\text{exp}}$ for an experimentally measured fingerprint or $g^{\text{A}}$ and $g^{\text{B}}$ to distinguish two different functions. }

% \NOTE{Note: Let's use parentheses () for function dependencies, e.g., $f(x)$ and square brackets for indicating computation order, e.g., $p[J-1]$.}

% \NOTE{Note: The table of contents will be removed later.}\\[4.pt]
% \tableofcontents
% \newpage
% % to be removed later >

% \clearpage

\begin{figure}[ht]
    \centering
    \includegraphics[width=0.8\textwidth]{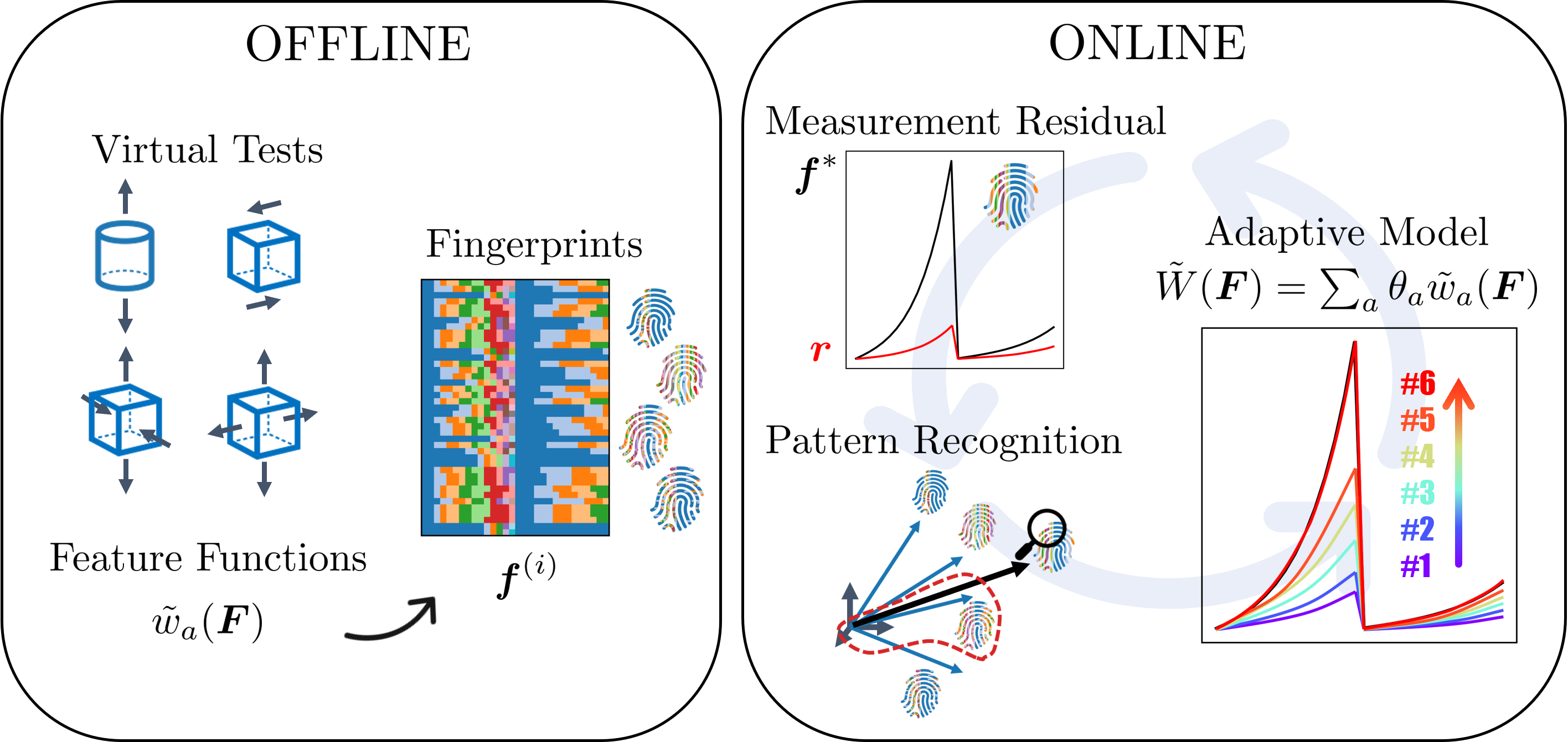}
    \caption{%
        Illustration of adaptive \MF{}.
        In the offline phase, the mechanical response of different modeling features is computed numerically for different standardized mechanical test, and saved as feature fingerprints in the database. 
        In the online phase, different modeling features are successively added to the material model to reconstruct the experimentally measured fingerprint.}
    \label{fig:abstract}
\end{figure}
% $\tilde w_a(\bfF)$

\section{Introduction}
\label{sec:introduction}

\MF{} is a novel method for the rapid discovery of material models \citep{flaschel_material_2026}.
It operates by precomputing a database of material responses, a.k.a. fingerprints, and their corresponding material models in an offline stage, which is subsequently leveraged in an online stage to identify the appropriate material model for an unseen experimental measurement via a pattern-recognition algorithm.
In its current form, however, \MF{} is limited to selecting only those material models that are explicitly represented in the database.
To overcome this limitation, we propose an adaptive \MF{} framework as summarized in \cref{fig:abstract}.
Adaptive \MF{} enables not only the selection of material models contained in the database, but also the discovery of arbitrary combinations thereof by successively adding terms from the database to the model.
In contrast to the original \MF{} method \citep{flaschel_material_2026}, this strategy facilitates the discovery of complex constitutive models with multiple modeling features, including the multi-term Ogden model \citep{ogden_large_1972} and the anisotropic Holzapfel-Gasser-Ogden model \citep{cowin_new_2004}.

Before describing \MF{} and its adaptive modification, we briefly place it in the context of existing methods for material characterization.
The classical strategy for material characterization is to assume a mathematical description of the material behavior that depends on a set of material parameters, which are then identified by minimizing the mismatch between model predictions and experimentally measured data \citep{gdoutos_mechanical_2024}.
In simple experimental setups, the strain state can be assumed to be homogeneous across the specimen.
Under this assumption, such experiments provide data pairs of strains and stresses that can be used to calibrate the material parameters.
However, because these simple experiments are limited to very basic deformation modes, they are often criticized for providing only sparse information in the high-dimensional strain space.
Consequently, experiments involving complexly shaped specimens have gained increasing popularity because they generate rich, heterogeneous strain states, which can be captured over the specimen surface using camera-based measurement systems \citep{grediac_principle_1989,hild_digital_2006}.
More specifically, the measured displacement fields and reaction forces can be used for material characterization by minimizing the mismatch between experimental data and data predicted by simulations of the corresponding boundary value problem, or by minimizing the residuals of the governing partial differential equations, see the overviews by \cite{avril_overview_2008,roux_optimal_2020,romer_reduced_2025}.
These methods can, however, be computationally demanding, particularly when they require repeated solutions of either the global boundary value problem or the local nonlinear equations at each material point and time step.

In contrast to a priori assuming a material model and calibrating its parameters, modern methods for material characterization impose fewer assumptions on the mathematical form of the material model, which promises enhanced fitting accuracy. For example, neural networks can be used to represent mechanical material behavior \citep{ghaboussi_knowledgebased_1991}.
A recent line of research is particularly concerned with designing such networks to satisfy the fundamental requirements imposed by physics \citep{asad_mechanics-informed_2022,klein_polyconvex_2022,klein_polyconvex_2025,masi_evolution_2023,rosenkranz_comparative_2023,linden_neural_2023,jadoon_automated_2024,kalina_neural_2025,flaschel_convex_2025,geuken_input_2025,holthusen2026complement}. Another line of research focuses on the discovery of concise and interpretable mathematical expressions for material models. Popular methods include symbolic regression \citep{abdusalamov_automatic_2023}, the EUCLID approach \citep{flaschel_unsupervised_2021,flaschel_automated_2023-2}, and CANNs \citep{linka_new_2023,linka_automated_2023-1,holthusen_theory_2024,holthusen2026generalized,martonova_automated_2024,martonova_generalized_2025}.
As with traditional material models, data-driven material models can be informed by stress-strain data pairs or by full-field displacement and reaction force data \citep{wang_inference_2021,thakolkaran_nn-euclid_2022,anton_physics-informed_2022,benady_unsupervised_2024,wiesheier_versatile_2024}.
Finally, in the extreme case, the formulation of the material model is omitted altogether, and simulations are directly informed by a database of stress-strain pairs, which serves as a surrogate for the material model \citep{kirchdoerfer_data-driven_2016,ibanez_data-driven_2017}.
For an overview of data-driven material modeling, we refer to \cite{fuhg_review_2024}.

Despite their enhanced fitting accuracy and broad applicability, all aforementioned data-driven material characterization methods share the requirement of solving an optimization problem.
Depending on the experimental setup and the underlying material modeling assumptions, this optimization problem can be computationally expensive, non-convex, and subject to nonlinear parameter constraints, such that convergence to a global optimum is often not guaranteed.
To address these challenges, we proposed the two-stage \MF{} approach.
In an offline stage, a database of mechanical material responses is generated for a range of material models and parameter sets by simulating a set of standardized experiments.
These responses can be interpreted as characteristic fingerprints of the corresponding material models.
Once generated and made publicly available, this database provides a reusable resource for rapid material model discovery.
When a new experiment is conducted on an uncharacterized material, its experimental fingerprint is extracted and compared against the database to discover the material models and parameter sets that best match the observed response.
In this way, material characterization becomes both computationally efficient, since no continuous optimization problem needs to be solved, and robust, as \MF{} guarantees identification of the best matching fingerprint contained in the database rather than convergence to a local optimum.
Following a numerical proof of concept \citep{flaschel_material_2026}, we validated \MF{} on experimental data \citep{martonova_material_2026}.
More recently, we demonstrated that \MF{} can be applied to data obtained from camera-based measurement systems \citep{flaschel_unsupervised_2026}, achieving speed-ups of several orders of magnitude compared to conventional approaches that rely on minimizing the mismatch between experimental data and predictions obtained from simulations of the corresponding boundary value problem.

In this work, we propose an adaptive strategy for \MF{} in which modeling features are successively incorporated into the material model to progressively improve the description of the experimental data.
As in the original approach, the adaptive \MF{} method is divided into an offline and an online stage.

\textbf{Offline:} Define a set of feature functions to describe the material model and generate a database of fingerprints for the modeling features, considering different material parameters.
% \begin{itemize}
%     \item[1:] Define a set of feature functions for describing the material model.
%     \item[2:] Generate a database of fingerprints for the modeling features considering different material parameters.
% \end{itemize}

\textbf{Online:}
% \begin{itemize}
%   \item[1:] Measure the experimental fingerprint and initialize the fingerprint predicted by the model to zero.
%   \item[2:] Compute the residual between the fingerprint predicted by the model and the experimental fingerprint.
%   \item[3:] Identify the database fingerprint closest to the residual and add the corresponding feature to the model. %with a weighting factor smaller than one.
%   \item[4:] Repeat Steps~2 and~3 a predefined number of times.
% \end{itemize}
\begin{itemize}
  \item[1:] Measure the experimental fingerprint and initialize the model prediction.
  \item[2:] Compute the residual between the model prediction and the experimental fingerprint.
  \item[3:] Select the database fingerprint that best matches the residual and update the model with the corresponding feature.
  \item[4:] Repeat Steps~2--3 for a fixed number of iterations.
\end{itemize}

We note that the proposed strategy shares conceptual similarities with sparse regression-based approaches for material model discovery \citep{flaschel_unsupervised_2021,mcculloch_automated_2024}. Traditional sparse regression methods begin with a large set of candidate features and employ $L_1$-regularization to enforce sparsity, thereby reducing the number of active features in the resulting model. In contrast, our adaptive strategy starts from a zero solution and incrementally adds features, producing a model that is sparse by construction, while allowing the user to specify the desired number of features. This approach shares similarities with forward stepwise regression \citep{hastie_elements_2009} and the LARS algorithm \citep{efron_least_2004,flaschel_non-smooth_2025,urreaquintero_automated_2026}. However, unlike our methods, these methods do not rely on a precomputed database. It is important to emphasize that, in all aforementioned methods, including our proposed approach, there is no guarantee of identifying the globally optimal combination of features from the database, which is a combinatorial problem that quickly becomes intractable as the number of potential features increases.

In this contribution, we focus on incompressible hyperelastic materials. We consider isotropic modeling features that depend on the principal stretches $\lambda_i$ ($i=1,\dots,3$) and on the principal area changes $\lambda_i \lambda_j$ ($i \neq j$), as well as anisotropic features that depend on the fiber stretch $\lambda_{\boldsymbol{a}}$ and the area change orthogonal to the fiber direction $A_{\boldsymbol{a}}$. However, alternative feature choices, for instance, based on the principal strain invariants \citep{hartmann_polyconvexity_2003}, could be incorporated in a straightforward manner. By considering ten distinct features in this work, the model ansatz is sufficiently expressive to capture several well-established material models from the literature, including the multi-term Ogden model \citep{ogden_large_1972} and the anisotropic Holzapfel-Gasser-Ogden model \citep{cowin_new_2004}. Furthermore, the chosen feature set allows for the direct incorporation of physical constraints, such as thermodynamic consistency, objectivity, and a stress-free reference configuration. We implement a simple switch that can be activated to enforce polyconvexity by restricting the database to an appropriate subset of entries. We establish databases for uniaxial tension/compression, pure shear, and biaxial tension/compression tests, and demonstrate the adaptive \MF{} approach by applying it to data obtained from isotropic rubber materials \citep{treloar_stress-strain_1944} as well as anisotropic animal skin \citep{linka_automated_2023-1}.

\section{Adaptive \MF{}}
The original \MF{} method \citep{flaschel_material_2026, flaschel_unsupervised_2026, martonova_material_2026} is restricted to discovering material models from an a priori defined lookup table of material models and associated parameters. This restriction limits the flexibility of the resulting models in describing experimental data. In this work, we therefore propose an adaptive strategy for \MF{} that enables not only the discovery of individual material models but also combinations of different material models through an iterative pattern recognition algorithm. This approach is expected to enhance fitting accuracy. In the following, we detail the experimental setups considered in this work, the definition of the fingerprints, the generation of the database, and the iterative pattern recognition algorithm underlying the adaptive \MF{} framework.

\subsection{Standardized experiments and fingerprint definition}
\label{sec:experiments}

\begin{figure}[ht]
    \centering
    \includegraphics[width=0.95\textwidth]{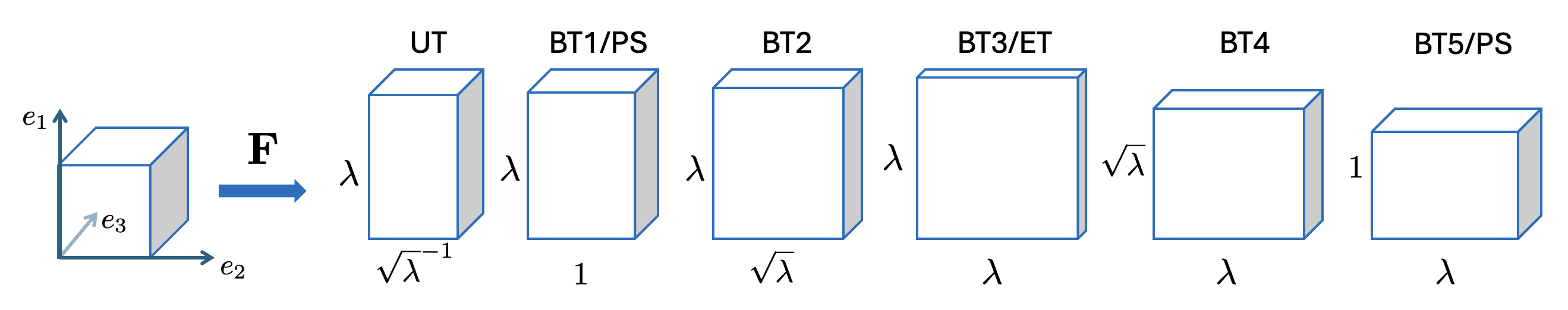}
    \caption{%
Stretch-controlled homogeneous deformation modes used for adaptive material fingerprinting under \CHANGE{the} incompressibility condition. All experiments are parameterized by a single stretch $\lambda$.  BT1 and BT5 correspond to pure shear experiments, while BT3 represents equibiaxial tension.}
    \label{fig:deformation_modes}
\end{figure}

\CHANGE{We consider six standardized homogeneous stretch-controlled experiments, namely uniaxial tension (UT) and five biaxial tension (BT1-BT5) experiments, under the assumption of incompressibility, i.e.,~$\det \mathbf{F}=1$. A schematic overview of the deformation modes is provided in \cref{fig:deformation_modes}. We focus on experiments in which $\bfF$ is diagonal and parametrize each experiment by a single control variable~$\lambda$. To this end, we set one diagonal component of $\bfF$ equal to the control variable $\lambda$, prescribe a relation between $\lambda$ and a second diagonal component, and determine the third diagonal component from the incompressibility condition.}

The corresponding deformation gradients are
\begin{equation}
\begin{aligned}
\text{UT:} \quad 
& \bfF = \mathrm{diag}\{\lambda, \lambda^{-1/2}, \lambda^{-1/2}\}
& \qquad
\text{BT3 (ET):} \quad 
& \bfF = \mathrm{diag}\{\lambda, \lambda, \lambda^{-2}\} \\[6pt]
\text{BT1 (PS):} \quad 
& \bfF = \mathrm{diag}\{\lambda, 1, \lambda^{-1}\}
& \qquad
\text{BT4:} \quad 
& \bfF = \mathrm{diag}\{\sqrt{\lambda}, \lambda, \lambda^{-3/2}\} \\[6pt]
\text{BT2:} \quad 
& \bfF = \mathrm{diag}\{\lambda, \sqrt{\lambda}, \lambda^{-3/2}\}
& \qquad
\text{BT5 (PS):} \quad 
& \bfF = \mathrm{diag}\{1, \lambda, \lambda^{-1}\}
\end{aligned}
\label{eq:load_cases}
\end{equation}

\CHANGE{We note that BT3 represents equibiaxial tension (ET), while BT1 and BT5 correspond to two distinct pure shear (PS) experiments. The latter are included to probe directional effects relevant for anisotropic behavior.}

To define the fingerprints, we compute the first Piola-Kirchhoff stress component $P_{11}$ for UT at $n_{\text{UT}} = 1000$ equidistant stretch levels, which span the interval $[0.1, 10]$.
For the five biaxial loading cases, we compute the stress components $P_{11}$ and $P_{22}$ for $n_{\text{BT}} = 1000$ values of the control variable in the interval $[0.1, 10]$.
All stress measurements are collected in a fingerprint vector $\bff \in \Rset^{n_f}$, where $n_f = n_{\text{UT}} + 2 \cdot 5 \cdot n_{\text{BT}}$.
Because different materials exhibit different stress responses in the same standardized experiments, the vector $\bff$ can be interpreted as the material's fingerprint that encodes the information of the material's mechanical behavior.
As we discuss later, to generate a database of fingerprints, we will numerically generate a number of $n_d$ fingerprints $\bff^{(i)}$ with $i=1,\dots,n_d$, by considering $n_d$ different material modeling features and parameter realizations.

% To define the fingerprints, we compute the first Piola-Kirchhoff stress component $P_{11}$ for UT, PS, and ET deformation modes at $n_{\text{UT}} = n_{\text{PS}} = n_{\text{ET}} = 1000$ equidistant stretch levels, which span the interval $[0.1, 10]$. For the simple shear experiment, we compute the stress component $P_{12}$ for $n_{\text{SS}} = 1000$ equidistant shear strain values $\gamma \in [0.001, 10]$. For the general biaxial experiments, we compute the stress components $P_{11}$ and $P_{22}$ for $n_{\text{BT}} = 1000$ values of the control variable in the interval $[0.1, 10]$ for all five biaxial loading cases.
% All stress measurements are collected in a fingerprint vector $\bff \in \Rset^{n_f}$, where $n_f = n_{\text{UT}} + n_{\text{SS}} + n_{\text{PS}} + n_{\text{ET}} + 2 \cdot 5 \cdot n_{\text{BT}}$.
% However, as the pure shear and equibiaxial experiments are special cases of the general biaxial experiment, the non-redundant information contained in the fingerprint amounts to $n_f = n_{\text{UT}} + n_{\text{SS}} + 2 \cdot 5 \cdot n_{\text{BT}}$.
% Because different materials exhibit different stress responses in the same standardized experiments, the vector $\bff$ can be interpreted as the material's fingerprint that encodes the information of the material's mechanical behavior.
% As we discuss later, to generate a database of fingerprints, we will numerically generate a number of $n_d$ fingerprints $\bff^{(i)}$ with $i=1,\dots,n_d$, by considering $n_d$ different material modeling features and parameter realizations.

\subsection{Adaptive fingerprint database}
\label{sec:adaptive fingerprint database}

In this work, we propose an adaptive fingerprint database that can be used to discover material model combinations using an iterative pattern recognition algorithm.
We assume incompressible hyperelasticity throughout this paper, such that the strain energy density $W$ can be written as
\begin{equation}
    W(\bfF,\bfa;\bftheta,\bfalpha) =
    \tilde W(\bfF,\bfa;\bftheta,\bfalpha)
    - p[\lambda_1\lambda_2\lambda_3 - 1],
    \label{eq:W_energy_incompressible}
\end{equation}
where $\bfF$ is the deformation gradient tensor, $\bfa$ is a normalized fiber direction, $p$ is a scalar-valued Lagrange multiplier, and $\lambda_i$ are the principal stretches, which are computed as the square roots of the eigenvalues of the right Cauchy-Green tensor $\bfC=\bfF^T\bfF$.
The constraint $\det(\bfF) = \lambda_1\lambda_2\lambda_3 = 1$ enforces incompressibility. \CHANGE{Under this constraint, only isochoric deformation states are admissible. Therefore, we will refer to $\tilde W$ as the isochoric strain energy density.}

In \cref{eq:W_energy_incompressible}, the material parameters $\bftheta$ and $\bfalpha$ are separated by a semicolon from the function arguments.
We distinguish between homogeneity parameters $\bftheta$ and non-homogeneity parameters $\bfalpha$. The homogeneity parameters fulfill the property $\tilde W(\bfF;c\bftheta,\bfalpha) = c \ \tilde W(\bfF;\bftheta,\bfalpha) \ \forall c \in \Rset$. This property is generally not fulfilled by the non-homogeneity parameters $\bfalpha$.

The isochoric strain energy density $\tilde W$ is constructed to satisfy the fundamental requirements of objectivity and material symmetry.
Assuming hyperelastic material behavior, the first Piola-Kirchhoff stress tensor follows from the differentiation of the strain energy density with respect to $\bfF$ as
\begin{equation}
    \bfP
    = \frac{\partial W}{\partial \bfF}
    = \frac{\partial \tilde{W}}{\partial \bfF}
    - p\,\mathrm{cof}(\bfF),
    \label{eq:stress}
\end{equation}
with $\text{cof}(\bfF) = \det (\bfF) \bfF^{-T}$.
A detailed discussion of the construction of physically admissible strain energy functions, including objectivity, material symmetry, and stress evaluation in terms of principal stretches, is provided in \cref{sec:theory}.

In adaptive \MF{}, we assume that the isochoric strain energy density is a linear combination of a number of $n_a$ feature functions $\tilde{w}_a$
\begin{equation}
    \label{eq:strain_energy_density_linear_combination}
    \tilde{W}(\bfF,\bfa;\bftheta,\bfalpha) = \sum_{a=1}^{n_a} \theta_a \ \tilde{w}_a(\bfF,\bfa;\bfalpha_i).
\end{equation}
The feature functions may take various forms, and multiple choices are possible. In this work, we consider feature functions defined as invariants of the deformation gradient tensor and the fiber direction.
We express these feature functions in terms of general scalar-valued functions $g : (0,\infty) \rightarrow \Rset$ and $h : (0,\infty) \rightarrow \Rset$ that may take various forms and are defined in more detail below. Specifically, we consider the two isotropic features
\begin{equation}
    \tilde{w}^{\text{I}}(\lambda_1,\lambda_2,\lambda_3;\bfalpha) = \sum_{j=1}^3 g(\lambda_j;\bfalpha)
\quad \mbox{and} \quad
    \tilde{w}^{\text{II}}(\lambda_1,\lambda_2,\lambda_3;\bfalpha) = \sum_{j=2}^3 \sum_{k<j} g(\lambda_j\lambda_k;\bfalpha),
    \label{eq:w12}
\end{equation}
which depend on the principal stretches $\lambda_j$ and the principal area changes $\lambda_j\lambda_k$.
\CHANGE{Similar modeling features have been used in the context of neural networks by \cite{st_pierre_principal-stretch-based_2023,vijayakumaran_consistent_2025,martonova_generalized_2025,tepole_polyconvex_2025}. Further, \cite{geuken_input_2025} have introduced neural network models that depend on the signed singular values of the deformation gradient, which are closely related to the principal stretches, and prove the universal approximation capability of these networks.}

In addition to the isotropic features, we consider anisotropic features.
We introduce a normalized fiber direction $\bfa = [\cos(\phi) \ \sin(\phi) \ 0]^T$ perpendicular to the $\bfe_3$-direction, where $\phi$ is the angle between the fiber and the $\bfe_1$-direction.
We note that the fiber direction is assumed to be aligned with the $\bfe_2$-direction throughout all experiments considered in this paper. We then consider the two anisotropic features
\begin{equation}
    \tilde{w}^{\text{III}}(\bfF,\bfa;\bfalpha) = h(\lambda_{\bfa};\bfalpha),
    \quad \mbox{and} \quad
    \tilde{w}^{\text{IV}}(\bfF,\bfa;\bfalpha) = h(A_{\bfa};\bfalpha),
    \label{eq:w34}
\end{equation}
where $\lambda_{\bfa} = \sqrt{\bfa^T \bfC \bfa}$ and
$A_{\bfa} = \sqrt{\bfa^T \text{cof}(\bfC) \bfa}$ with $\text{cof}(\bfC) = \det (\bfC) \bfC^{-T}$ describe the stretch parallel to and the area change perpendicular to the fiber direction.
\CHANGE{These deformation measures are directly related to the mixed invariants of the right Cauchy-Green tensor $\bfC$ and the structural tensor $\bfM = \bfa \otimes \bfa$. Specifically, it is $\lambda_{\bfa} = \sqrt{I_4}$ and $A_{\bfa} = \sqrt{I_5}$, where $I_4 = \bfC : \bfM$ and $I_5 = \text{cof}(\bfC) : \bfM$.}

\begin{figure}[!h]
    \centering
    \begin{subfigure}[b]{0.33\textwidth}
        \centering
        \includegraphics[width=\textwidth]{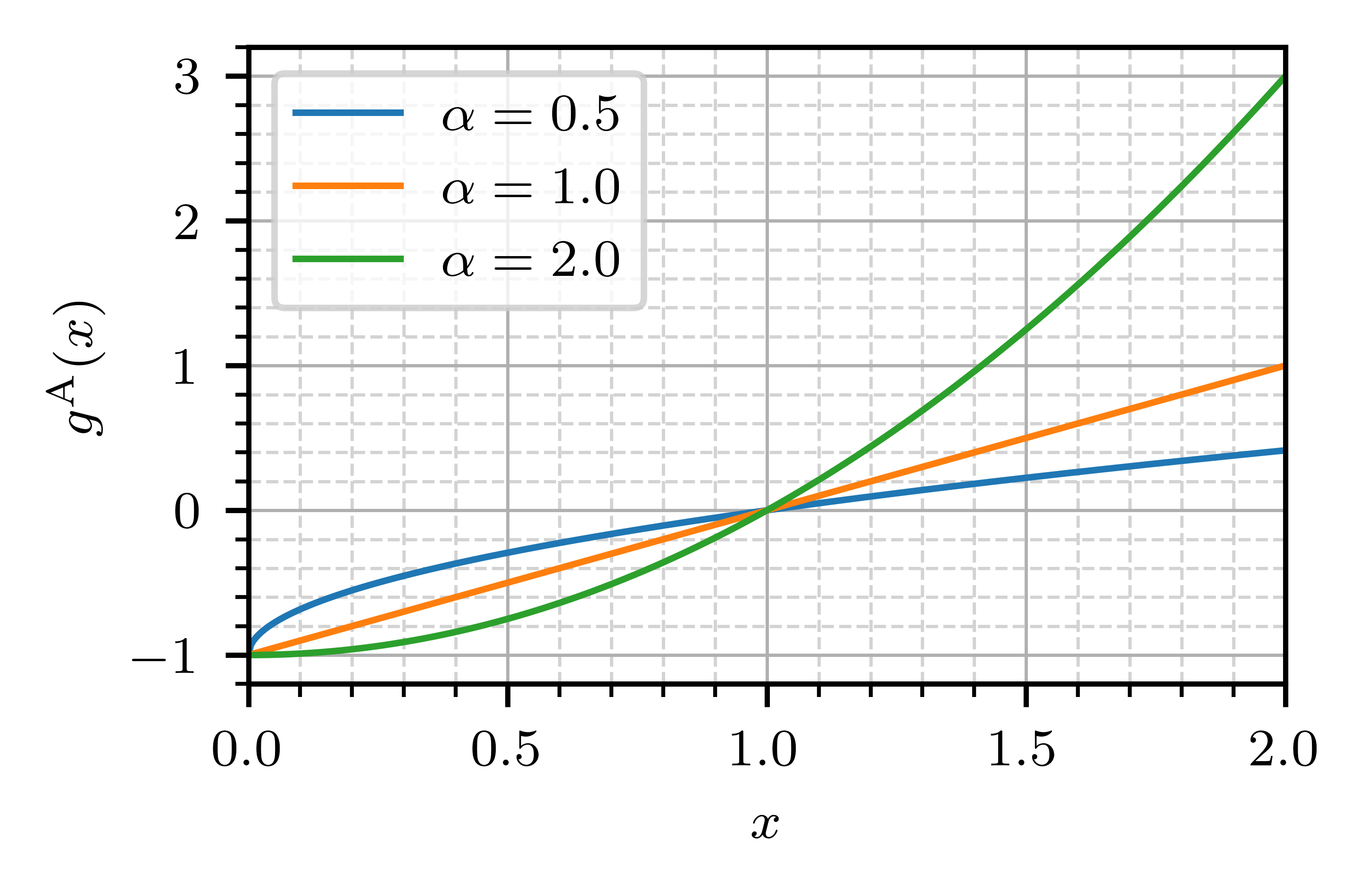}
        \caption{$g^{\text{A}}(x)$.}
    \end{subfigure}
    \begin{subfigure}[b]{0.33\textwidth}
        \centering
        \includegraphics[width=\textwidth]{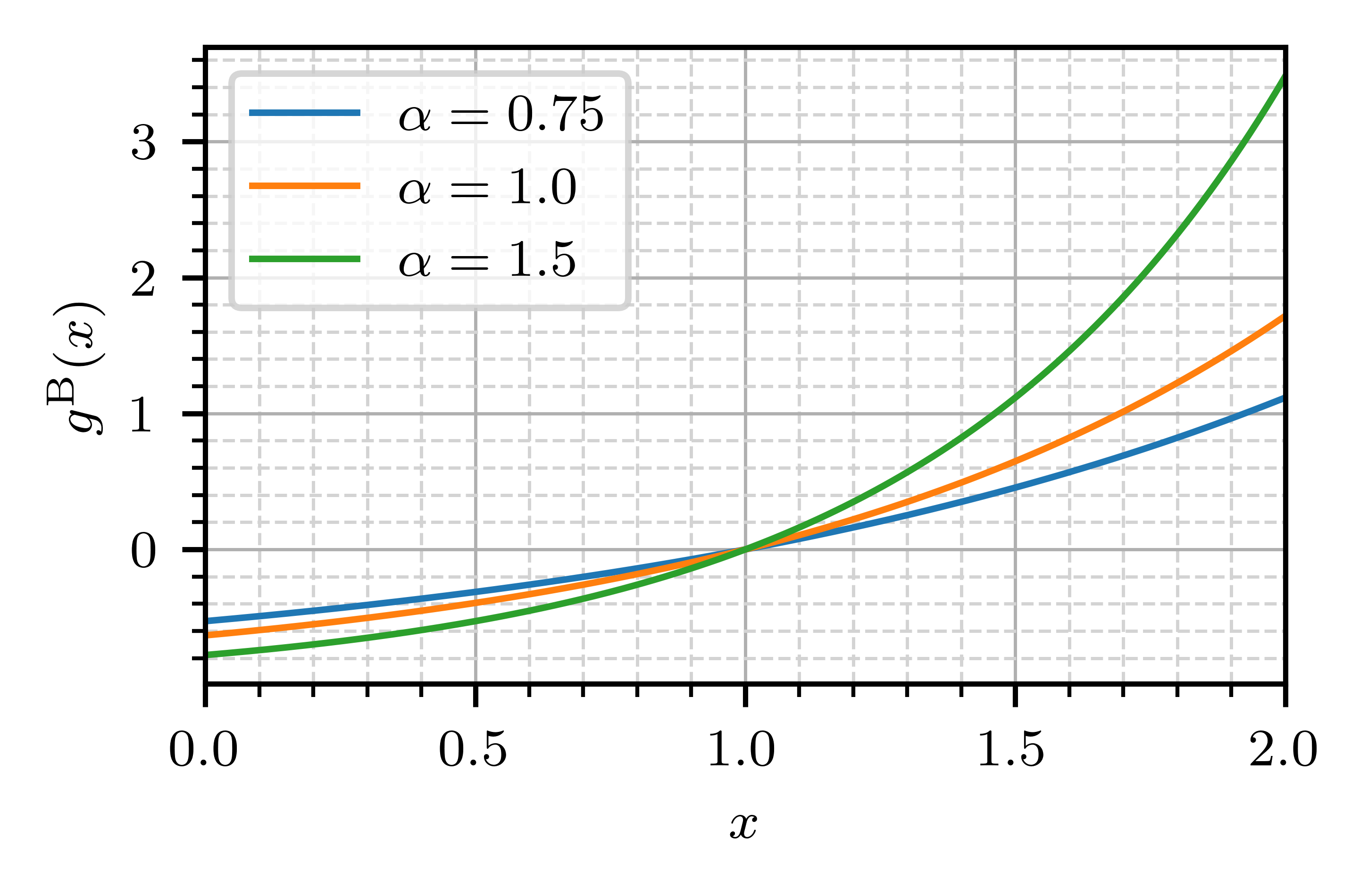}
        \caption{$g^{\text{B}}(x)$.}
    \end{subfigure}
    \begin{subfigure}[b]{0.33\textwidth}
        \centering
        \includegraphics[width=\textwidth]{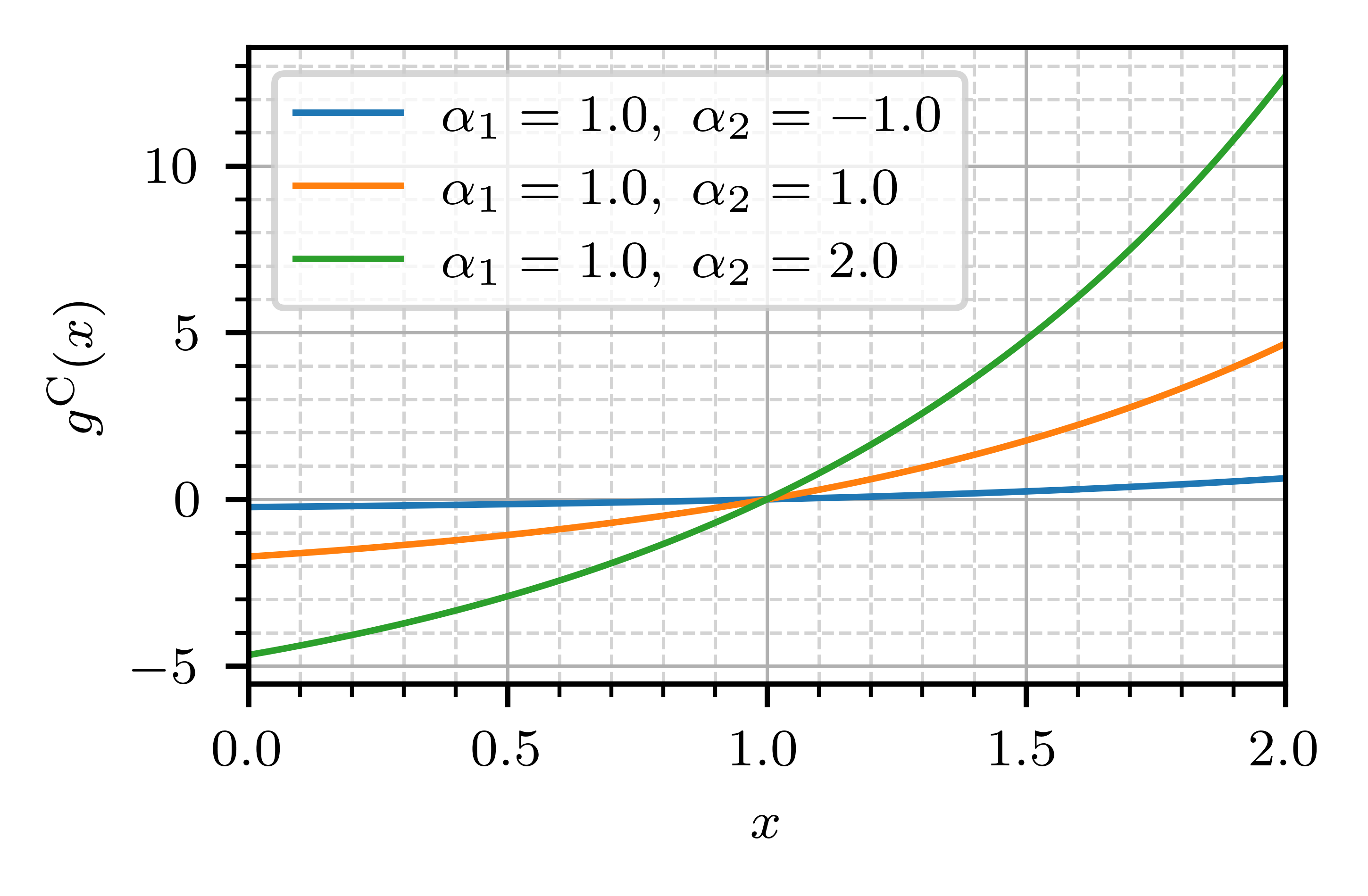}
        \caption{$g^{\text{C}}(x)$.}
    \end{subfigure}
    \caption{Functions considered for the construction of isotropic strain energy density features \CHANGE{depending on the principal stretches $\lambda_j$ or the principal area changes $\lambda_j\lambda_k$.}}
    \label{fig:functions_g}
\end{figure}

\begin{figure}[!h]
    \centering
    \begin{subfigure}[b]{0.33\textwidth}
        \centering
        \includegraphics[width=\textwidth]{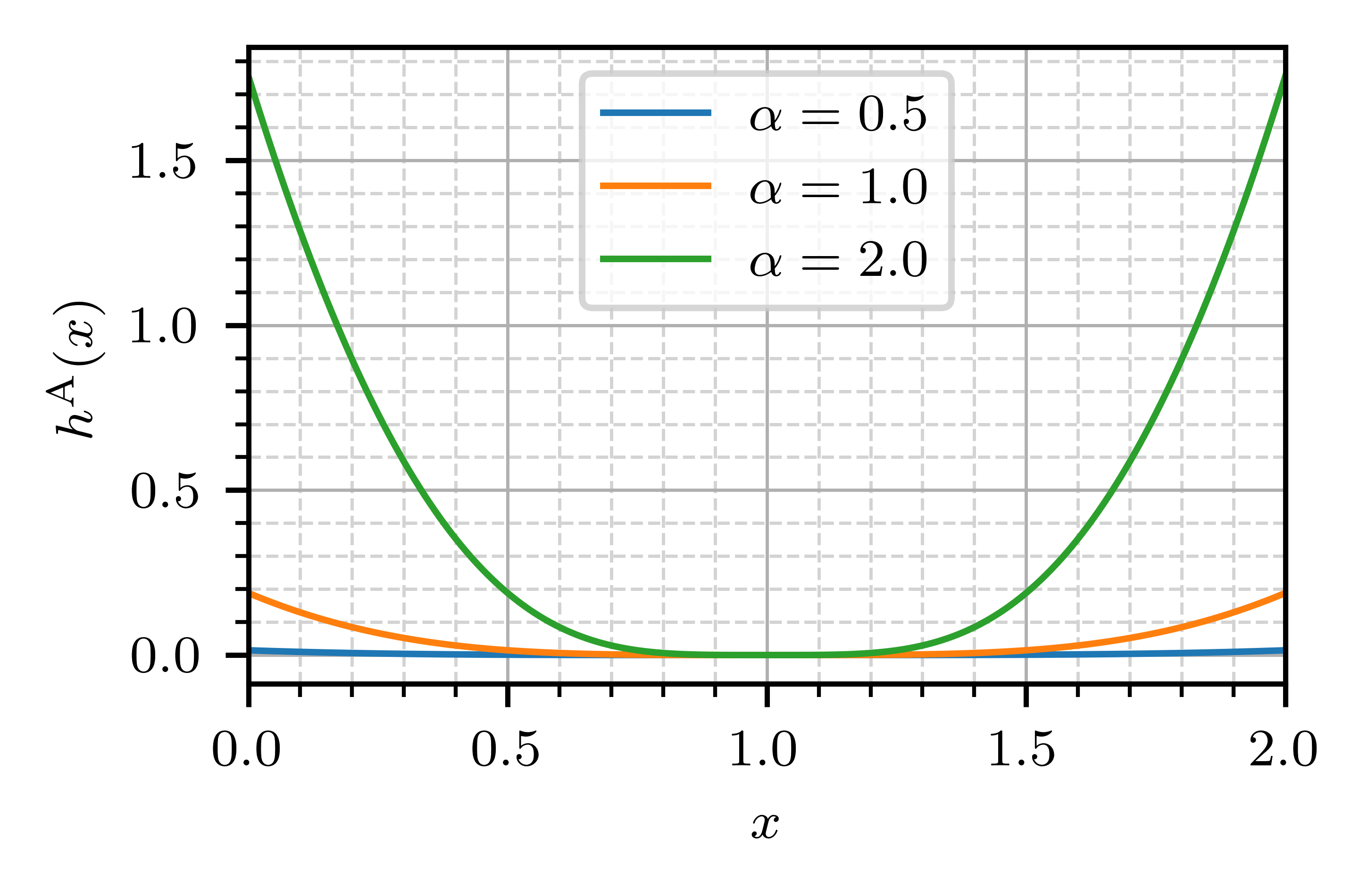}
        \caption{$h^{\text{A}}(x)$.}
    \end{subfigure}
    \begin{subfigure}[b]{0.33\textwidth}
        \centering
        \includegraphics[width=\textwidth]{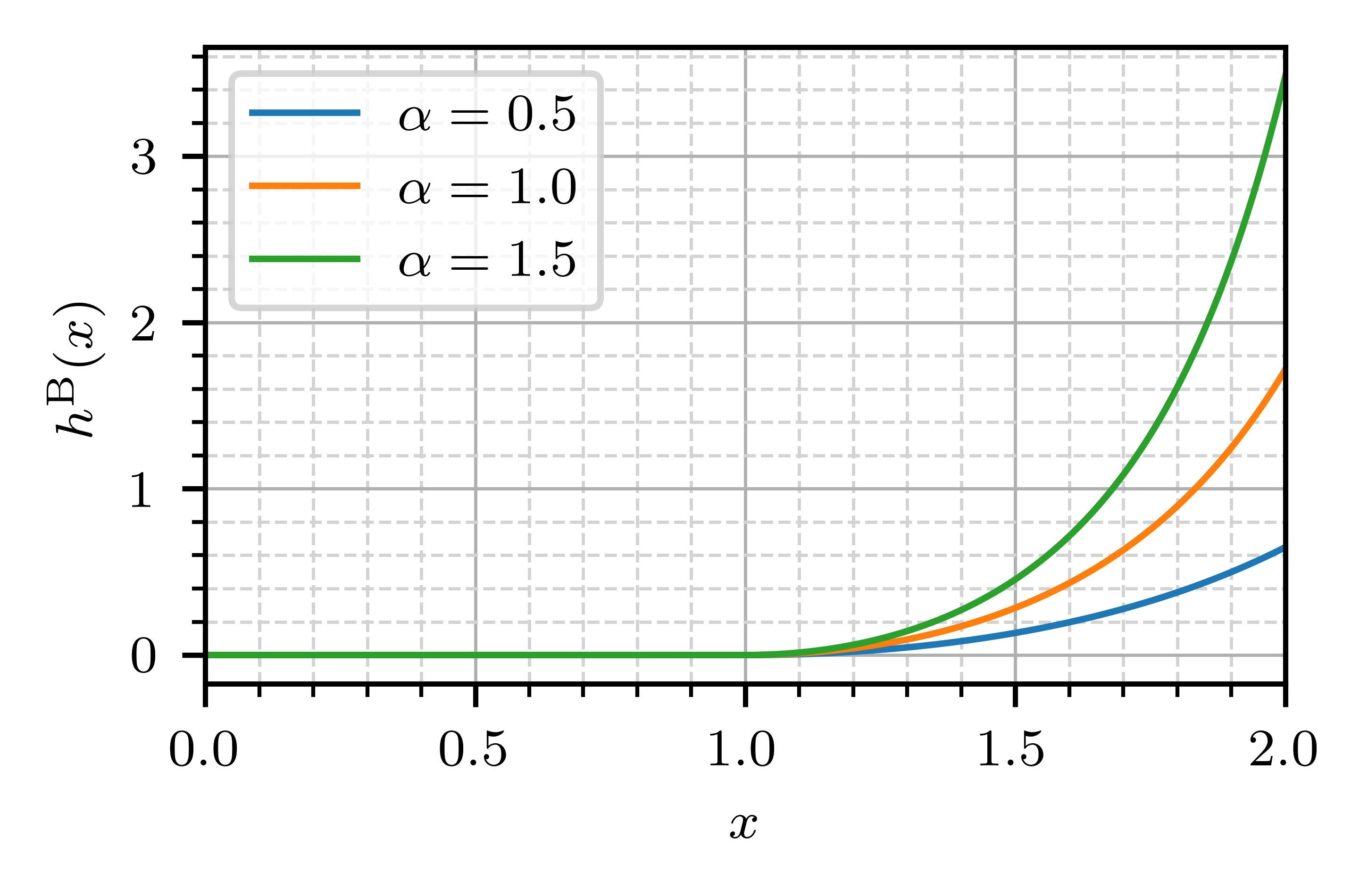}
        \caption{$h^{\text{B}}(x)$.}
    \end{subfigure}
    % \begin{subfigure}[b]{0.33\textwidth}
    %     \centering
    %     \includegraphics[width=\textwidth]{fig/function_hC.png}
    %     \caption{$h^{\text{C}}(x)$.}
    % \end{subfigure}
    \caption{Functions considered for the construction of anisotropic strain energy density features \CHANGE{depending on the fiber stretch $\lambda_{\bfa}$ or the area change perpendicular to the fiber direction $A_{\bfa}$.}}
    \label{fig:functions_h}
\end{figure}

We note that, while this work is restricted to the feature functions $\tilde{w}^{\text{I}}$ to $\tilde{w}^{\text{IV}}$, the consideration of alternative feature functions in the future is equally possible. The functions $g$ are assumed to take one of the following forms
\begin{equation}
    g^{\text{A}}(x) = x^{\alpha} - 1
   \qquad \mbox{or} \qquad
    g^{\text{B}}(x) = \exp{(\alpha[x-1])} - 1
   \qquad \mbox{or} \qquad
    g^{\text{C}}(x) = \exp{(\alpha_1[x-1] + \alpha_2)} - \exp{(\alpha_2)}.
    \label{eq:gABC}
\end{equation}
Further, the functions $h$ are assumed to take one of the following forms
\begin{equation}
    h^{\text{A}}(x) = \log (\cosh (\alpha[x-1]))^2
   \qquad \mbox{or} \qquad
    h^{\text{B}}(x) = \exp ( \alpha [\max (\{x-1,0\})]^2) - 1.
    \label{eq:hAB}
\end{equation}
The functions $g$ and $h$ are illustrated for different values of $\alpha$ in \cref{fig:functions_g,fig:functions_h}.
These functions are chosen to provide a flexible ansatz for the strain energy density function while encompassing several well-established material models from the literature.
For example, combining the isotropic feature $\tilde{w}^{\text{I}}$ with the function $g^{\text{A}}$ yields the Ogden model \citep{ogden_large_1972}, while combining multiple terms with different values of $\alpha$ naturally leads to the multi-term Ogden model.
Furthermore, combining the anisotropic feature $\tilde{w}^{\text{III}}$ with the function $h^{\text{B}}$ results in the Holzapfel-Gasser-Ogden model \citep{cowin_new_2004}. \CHANGE{We note that the function $h^{\text{B}}(x)$ is not twice continuously differentiable at $x=1$. If a twice continuously differentiable strain energy density function is required within the proposed framework, all features associated with $h^{\text{B}}$ can either be excluded during the pattern recognition stage, or the function can be replaced by a suitably smooth surrogate.}

\CHANGE{We briefly comment on the stress normalization condition for elastic materials, i.e., $\bfP=\mathbf{0}$ for $\bfF=\bfI$.  
Using the chain rule, the Piola--Kirchhoff stress is given by
\begin{equation}
    \bfP =
    \sum_{a=1}^{n_a}
    \theta_a
    \frac{\partial \tilde{w}_a}{\partial \mathcal{I}}
    \frac{\partial \mathcal{I}}{\partial \bfF}
    - p\,\mathrm{cof}(\bfF),
\end{equation}
where $\mathcal{I}$ denotes the invariants used in the feature functions, e.g., principal or fiber stretches.
Evaluating the derivatives $\partial \mathcal{I}/\partial \bfF$ at the undeformed configuration shows that they are generally nonzero, cf.~\ref{sec:app_stress_calculation}. For isotropic features, however, these derivatives reduce to the identity tensor for $\bfF=\bfI$. Together with $p\,\mathrm{cof}(\bfF)=p\,\bfI$, the hydrostatic pressure term therefore naturally satisfies the stress normalization condition for incompressible isotropic behavior.
For anisotropic features, in contrast, the derivatives $\partial \mathcal{I}/\partial \bfF$ do not reduce to the identity tensor. Hence, the isotropic pressure term alone cannot enforce $\bfP=\mathbf{0}$ in the undeformed configuration. To address this issue, we require the first derivatives of the anisotropic feature functions to vanish at $\bfF=\bfI$. This additional condition motivates the different design of the anisotropic features $h$ compared to the isotropic features $g$.
}

Within the adaptive \MF{} framework, polyconvexity (\cite{ball_convexity_1976}) can be enforced by restricting the admissible feature functions and their parameter ranges, see \cref{sec:theory} for a detailed discussion.
This enables us to control, a priori, whether the resulting strain energy density satisfies a sufficient condition for the existence of minimizers when combined with an appropriate growth condition.
In particular, polyconvexity is ensured if the functions $g$ and $h$ in \cref{eq:gABC}-\cref{eq:hAB} are convex and monotone, which leads to simple parameter constraints: $g^A$ requires $\alpha \geq 1$, $g^B$ requires $\alpha \geq 0$, and $g^C$ requires $\alpha_1 \geq 0$, while $h^B$ is admissible for $\alpha \geq 0$ and $h^A$ is excluded.
During database generation, each feature configuration is flagged according to whether it guarantees polyconvexity, allowing the user to optionally enforce this constraint in the online phase by restricting the selection to admissible features.

% The model in \cref{eq:strain_energy_density_linear_combination} can be constructed from ten different feature functions, which yields a total of $\sum_{k=1}^{n_a} \binom{10}{k}$ distinct models.
% For example, choosing $n_a=10$ yields more than a thousand possible models.

% We note that some of the considered modeling features require the evaluation of the exponential function, which can lead to overflow errors for large deformations. To circumvent this, we approximate the features by a second order Taylor-series around the point at which the slope of the strain energy reaches a predefined threshold, see \cref{sec:taylor} for details.
 
\begin{table}[h!]
\caption{Feature functions and material parameters considered during adaptive database generation. For all four feature functions $\tilde{w}^{\text{I}}$ to $\tilde{w}^{\text{IV}}$, we generate fingerprints considering all functions $g$ and $h$, respectively.}
\label{tab:database_supervised}
\centering
\begin{tabular}{|l|l|l|r|}
\hline
Function & Formula & Parameters ranges & \# Fingerprints \\ \hline

$g^{\text{A}}(x)$ & $x^{\alpha} - 1$ & $\alpha \in [-25.0,25.0]$ & 500 \\ 

$g^{\text{B}}(x)$ & $\exp{(\alpha[x-1])} - 1$ & $\alpha \in [0.1,25.0]$ & 500 \\

$g^{\text{C}}(x)$ & $\exp{(\alpha_1[x-1] + \alpha_2)} - \exp{(\alpha_2)}$ & $\alpha_1 \in [0.1,25.0]$, $\alpha_2 \in [-25.0,25.0]$ & 400 \\

$h^{\text{A}}(x)$ & $\log (\cosh (\alpha[x-1]))^2$ & $\alpha \in [0.1,25.0]$ & 500 \\ 

$h^{\text{B}}(x)$ & $\exp ( \alpha \max (x-1,0)^2) - 1$ & $\alpha \in [0.1,25.0]$ & 500 \\ 

\hline
\multicolumn{3}{|r|}{Isotropic database} & $n_d = 2 \times 1400 = 2800$ \\
\hline
\multicolumn{3}{|r|}{Anisotropic database} & $n_d = 2 \times 2400 = 4800$ \\
\hline
\end{tabular}
\end{table}

In this work, we generate two different databases, an isotropic database and an anisotropic database. For the isotropic database, we consider only isotropic features and, following our previous work \citep{martonova_material_2026}, generate fingerprints for uniaxial tension/compression (UT), pure shear (BT1), and equibiaxial tension/compression tests (BT3), \CHANGE{which are sufficient due to the direction-independent response.}
% For the anisotropic database, we consider both isotropic and anisotropic features and generate fingerprints for the biaxial tension/compression experiments with different stretch ratios $\lambda_1/\lambda_2$ as defined in \cref{sec:experiments}.
For the anisotropic database, we consider both isotropic and anisotropic features and generate fingerprints for the biaxial tension/compression experiments BT1-BT5, \CHANGE{where the inclusion of additional loading paths enables the characterization of directional effects. We compute the corresponding stress responses as described in \ref{sec:app_stress_calculation}.}
\cref{tab:database_supervised} summarizes the parameter ranges assumed during database generation, as well as the number of distinct fingerprints generated for each function.
Since the functions $g$ are employed for both features $\tilde{w}^{\text{I}}$ and $\tilde{w}^{\text{II}}$, and the functions $h$ for both features $\tilde{w}^{\text{III}}$ and $\tilde{w}^{\text{IV}}$, the isotropic and anisotropic databases comprise a total number of $2800$ and $4800$ individual functions and parameter choices, respectively.
This means that there are a total of $\sum_{k=1}^{n_a} \binom{2800}{k}$ or $\sum_{k=1}^{n_a} \binom{4800}{k}$ distinct combinations that could potentially be selected.
After being generated once, the total number of $n_d$ material fingerprints $\bff^{(i)}\in\Rset^{n_f}$ with $i=1,\dots,n_d$ is stored in a database and can be repeatedly used for fast material discovery using the adaptive pattern recognition algorithm described next.

\subsection{Experimental fingerprint acquisition}

In general, the deformation applied during the experiments is different from the deformation applied during database generation.
The values of the control variables assumed during database generation might not exactly match the experimentally applied control.
Therefore, we have to preprocess the experimental data before the iterative pattern recognition algorithm can be applied.
During preprocessing, the experimental measurements are brought into correspondence with the fingerprint database by mapping them onto identical deformation states.
The experimental dataset can comprise uniaxial tension or compression, pure shear, or biaxial tension or compression, as well as any combination of these loading modes evaluated at prescribed values of the control variables.
For consistency with the database, the experimental stress responses are estimated at the stretch or strain values stored in the fingerprint database using linear interpolation.
Whenever a database deformation state exceeds the range covered by the experimental measurements, the associated stress value is assigned as zero. As a result, this process produces a measured fingerprint vector $\bff^* \in \mathbb{R}^{n_f}$, which is sparse in regions where experimental data are unavailable.
We then define the masked experimental fingerprint $\hat\bff^* \in \Rset^{\hat n_f}$ and masked database fingerprints $\hat\bff^{(i)} \in \Rset^{\hat n_f}$ with $\hat n_f \leq n_f$, which contain only those entries of $\bff^*$ and $\bff^{(i)}$ at which $\bff^*$ is not zero.
Finally, we normalize the database fingerprints $\bar\bff^{(i)} = \hat\bff^{(i)}/\|\hat\bff^{(i)}\|$ and the corresponding homogeneity parameters $\bar\bftheta^{(i)} = \bftheta^{(i)}/\|\hat\bff^{(i)}\|$, as well as the experimental fingerprint $\bar\bff^* = \hat\bff^*/\|\hat\bff^*\|$.
% \be
% \label{eq:normalization}
% \bar\bff^{(i)} = \frac{\hat\bff^{(i)}}{\|\hat\bff^{(i)}\|},
% \ \bar\bftheta^{(i)} = \frac{\bftheta^{(i)}}{\|\hat\bff^{(i)}\|},
% \ \bar\bff^* = \frac{\hat\bff^*}{\|\hat\bff^*\|}.  
% \ee
Through the interpolation and masking procedure, we guarantee that the experimental fingerprint $\hat\bff^*$ and the database fingerprints $\hat\bff^{(i)}$ have identical lengths and contain stress values corresponding to the same values of the control variable.

\subsection{Adaptive pattern recognition algorithm}

In this contribution, we propose an iterative framework for discovering adaptive strain energy density functions with a tunable number of feature functions and material parameters.
The approach starts from a vanishing isochoric strain energy density function $\tilde W = 0$, and progressively enriches it by adding terms of the form $\theta_a \tilde{w}_a$ where the total number of terms is denoted by $n_a$.
Using this strategy, we build an adaptive strain energy density function that exhibits increasing fitting accuracy as more features are added to the function. The algorithm is detailed in \cref{alg:adaptive} and described in the following. We note that the online stage of adaptive \MF{} exhibits methodological parallels to forward stepwise regression \citep{hastie_elements_2009} and the LARS algorithm \citep{efron_least_2004,flaschel_non-smooth_2025,urreaquintero_automated_2026}. However, adaptive \MF{} differs in its utilization of a precomputed database and in its use of distinct step-size rules.

\begin{algorithm}
\caption{Adaptive Material Fingerprinting}\label{alg:adaptive}
\begin{algorithmic}
\State Given database fingerprints $\bar\bff^{(i)}$
\State Set $\tilde W=0$
\State Choose the maximum number of steps $n_a$ and the scaling parameter $s$
\For{$a=1,\dots,n_a$}
\State Compute masked fingerprint $\hat\bff$ dependent on the isochoric strain energy density $\tilde W$
\State $\hat \bfr  \gets \hat\bff^* - \hat\bff$, $\bar\bfr \gets \hat{\bfr} / \|\hat{\bfr}\|$
\State $i^* = \argmax_{i=1, \dots ,n_d} \ \bar\bff^{(i)} \cdot \bar\bfr$
\If{$a = n_a$}
    \State $s \gets 1$
\EndIf  
\State $\theta_a = s \ \|\hat\bfr\| \ \bar\bftheta^{(i^*)}$
\State $\tilde{w}_a = \tilde{w}^{(i)}$
\State $\tilde W \gets \tilde W + \theta_a\tilde{w}_a$
\EndFor
\State Compute $\hat\bff$ dependent on $\tilde W$
\For{$a=1,\dots,n_a$}
\State $\theta_a \gets \theta_a \ \|\hat\bff^*\| / \|\hat\bff\|$
\EndFor
\State $\tilde W \gets \sum_a \theta_a\tilde{w}_a$
\end{algorithmic}
\end{algorithm}

In each step $a$, we compute the fingerprint $\hat\bff\in\Rset^{\hat n_f}$ at the masked stretched values corresponding to the current isochoric strain energy density function $\tilde W$. 
We then define the residual between the current fingerprint and the experimental fingerprint $\hat\bfr = \hat\bff^* - \hat\bff$ and normalize it, $\bar\bfr = \hat{\bfr} / \|\hat{\bfr}\|$. Using a pattern recognition approach, we identify the fingerprint in the database that best represents the current residual. Specifically, this is achieved by selecting the database fingerprint that is closest to the residual according to
\be
\label{eq:pattern_recognition_supervised}
i^* = \argmax_{i=1, \dots ,n_d} \ \bar\bff^{(i)} \cdot \bar\bfr.
\ee
The feature corresponding to the identified fingerprint $\tilde{w}_a=\tilde{w}^{(i^*)}$ is next added to the current isochoric strain energy density function. If $a<n_f$, we set the corresponding homogeneity parameter $\theta_a$ to 
\be
\theta_a = s \ \|\hat\bfr\| \ \bar\bftheta^{(i^*)},
\ee
where $s \in (0,1]$ is a scalar hyperparameter that determines how much the residual should be reduced in each step. After adding $\theta_a \tilde{w}_a$ to the isochoric strain energy density function $\tilde W$, we proceed with the next step of the iterative procedure. In the final step, when $a=n_a$, we set $\theta_a = \| \hat\bfr \| \ \bar\bftheta^{(i^*)}$.
Finally, after the completion of the iterative procedure, we apply a final rescaling of all homogeneity parameters $\theta_a \ \|\hat\bff^*\| / \|\hat\bff\|$, where $\hat\bff$ is the masked fingerprint of the discovered material model.

The proposed \cref{alg:adaptive} depends on two user-defined hyperparameters, $n_a$ and $s$. The parameter $n_a$ determines the number of terms included in the discovered strain energy density function and thus governs the sparsity and interpretability of the resulting model. The scaling parameter $s$ controls the magnitude by which the strain energy density function is updated at each iteration. Both hyperparameters influence the fitting accuracy of the method. Their roles and appropriate selection are discussed at several points throughout this paper.

\section{Results}
\label{sec:results}

In the following, we apply the adaptive \MF{} to experimental data. To assess the goodness of fit, we use the coefficient of determination $R^2$ for each individual deformation mode, as well as the mean coefficient of determination across all considered deformation modes, called average $R^2$ in the following, see, e.g., \cite{martonova_material_2026} for the definition.
The performance of the adaptive \MF{} framework depends on the selection of the hyperparameters $n_a$ and $s$. In our experiments, we found that a heuristic choice of $n_a = 10$ and $s = 0.5$ consistently yields models with high fitting accuracy, see \cref{sec:heuristic_hyperparameters}. Therefore, when the primary objective is to maximize predictive performance and the number of terms in the model is of secondary importance, this parameter setting provides a reliable default. However, as we demonstrate in the following sections, these hyperparameters can be further optimized to produce sparse and more interpretable models while maintaining a high level of fitting accuracy. Consequently, we adopt this strategy in the remainder of this work.

\subsection{Isotropic rubber material}

To experimentally validate the proposed adaptive \MF{}, we apply it to experimental data of vulcanized rubber reported by \cite{treloar_stress-strain_1944} and listed in \cref{app: experimental data}.
Specifically, we consider uniaxial tension (UT), pure shear (BT1), and equibiaxial tension (BT3) tests conducted at temperatures of $20^{\circ}\mathrm{C}$ and $50^{\circ}\mathrm{C}$.
These datasets have previously been analyzed within the \MF{} framework in \cite{martonova_material_2026}, thereby providing a suitable benchmark for comparison.
Since the investigated materials do not exhibit a preferred direction, it is not appropriate to assume any specific fiber orientation. Consequently, the computation of the invariants $\lambda_{\bfa}$ and $A_{\bfa}$ is not meaningful. We therefore restrict the analysis to the isotropic feature database introduced in \cref{sec:adaptive fingerprint database}.

\begin{figure}[!ht]
    \centering
    \begin{subfigure}[b]{0.4\textwidth}
        \centering
        \includegraphics[width=\textwidth]{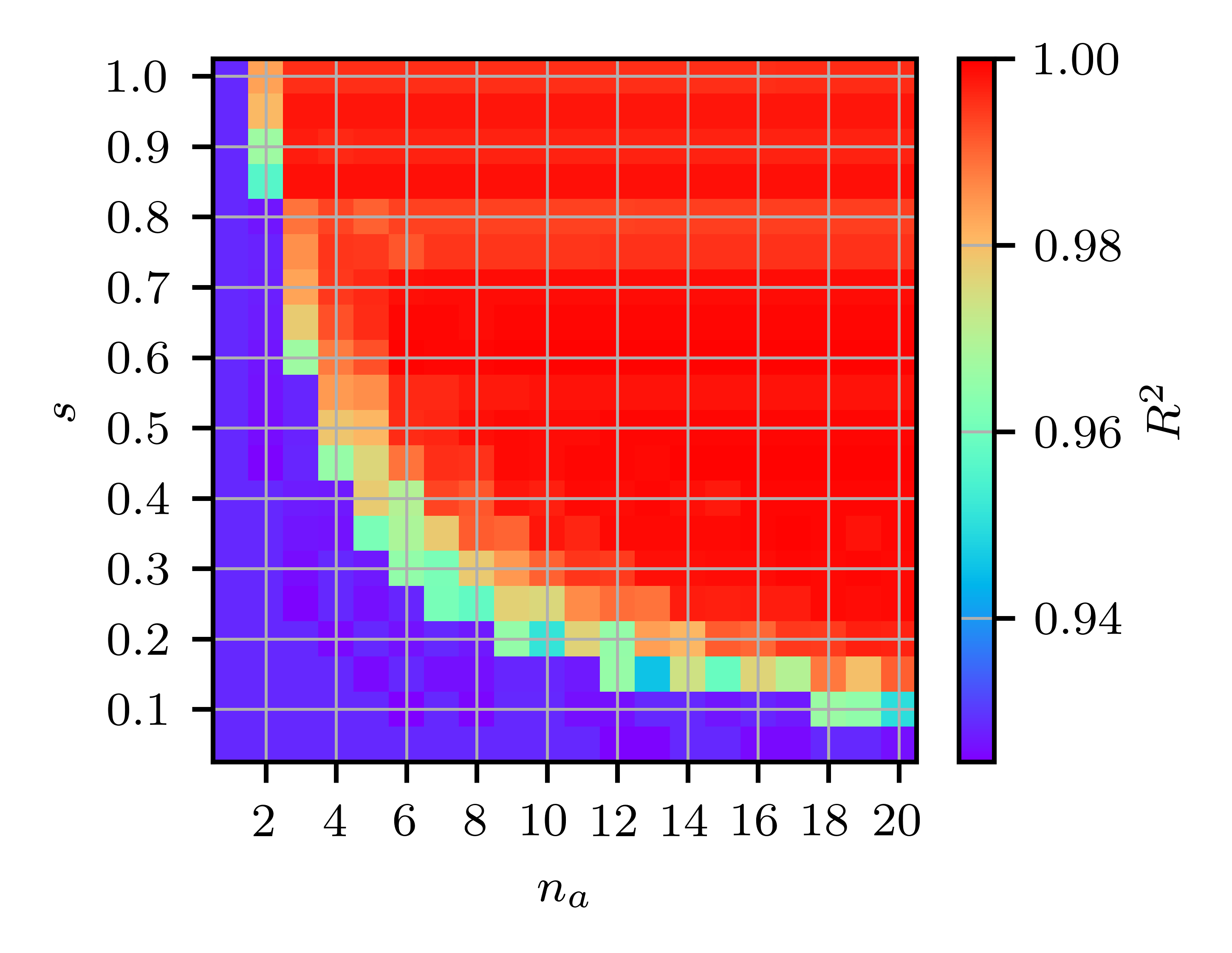}
        \caption{Rubber at $20^{\circ}\mathrm{C}$.}
    \end{subfigure}
    \begin{subfigure}[b]{0.4\textwidth}
        \centering
        \includegraphics[width=\textwidth]{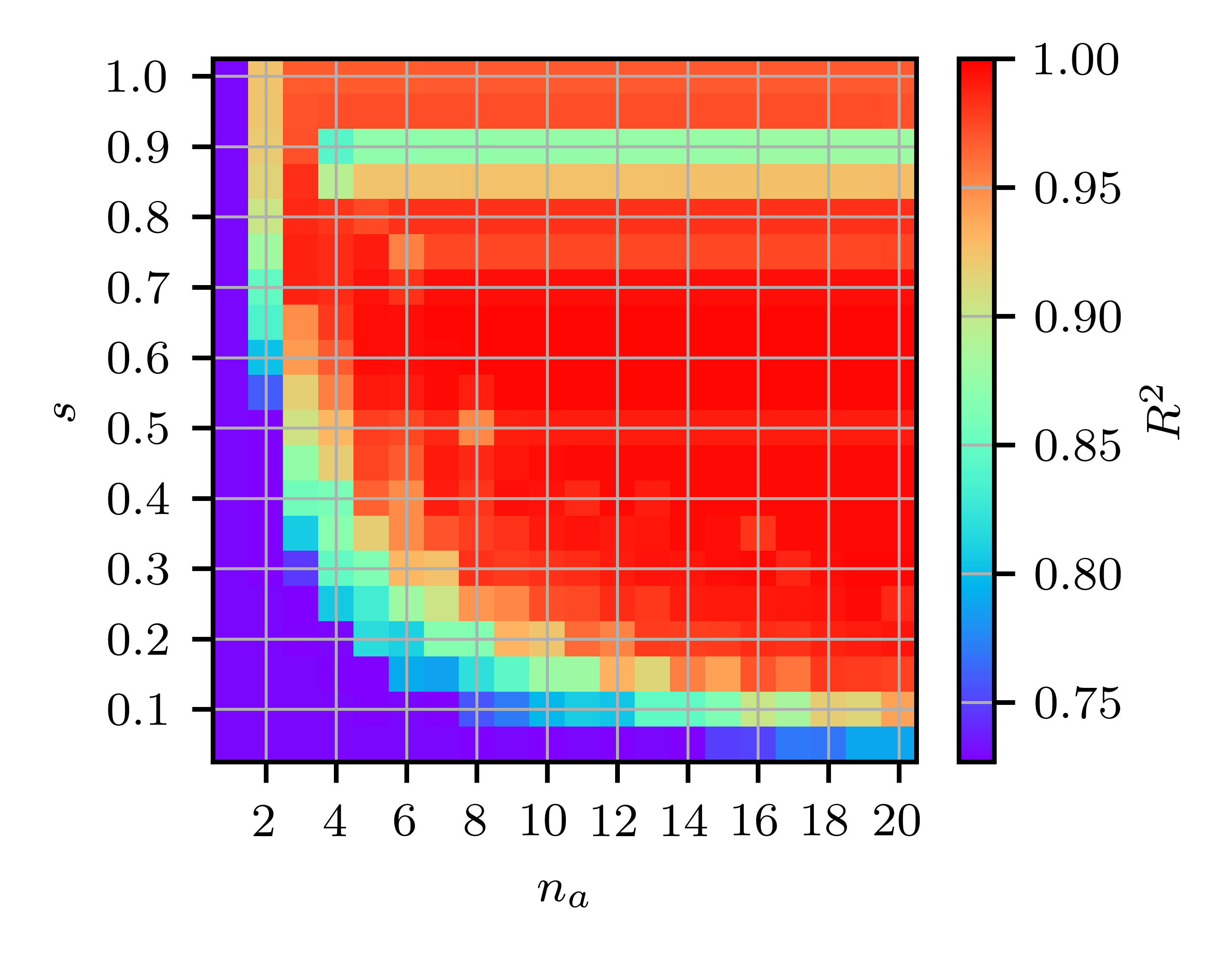}
        \caption{Rubber at $50^{\circ}\mathrm{C}$.}
    \end{subfigure}
    \caption{
    Hyperparameter analysis for adaptive \MF{} applied to rubber data.
    }
    \label{fig:rubber_hyperparameter_plot}
\end{figure}

As stated above, we first conduct a hyperparameter analysis for both datasets. To this end, we apply the adaptive \MF{} for different choices of the two hyperparameters. The resulting average $R^2$ values are illustrated in \cref{fig:rubber_hyperparameter_plot}. These figures can be used to select optimal values for the hyperparameters. As expected, we observe that the fitting accuracy tends to increase as the number of terms $n_a$ in the material model increases. We also observe that higher values of $s$ seem to be preferable for the two experimental datasets. However, as we will show later, this does not mean that higher values of $s$ are preferred in adaptive \MF{} in general.

\begin{figure}[!ht]
\centering
\includegraphics[width=0.75\textwidth]{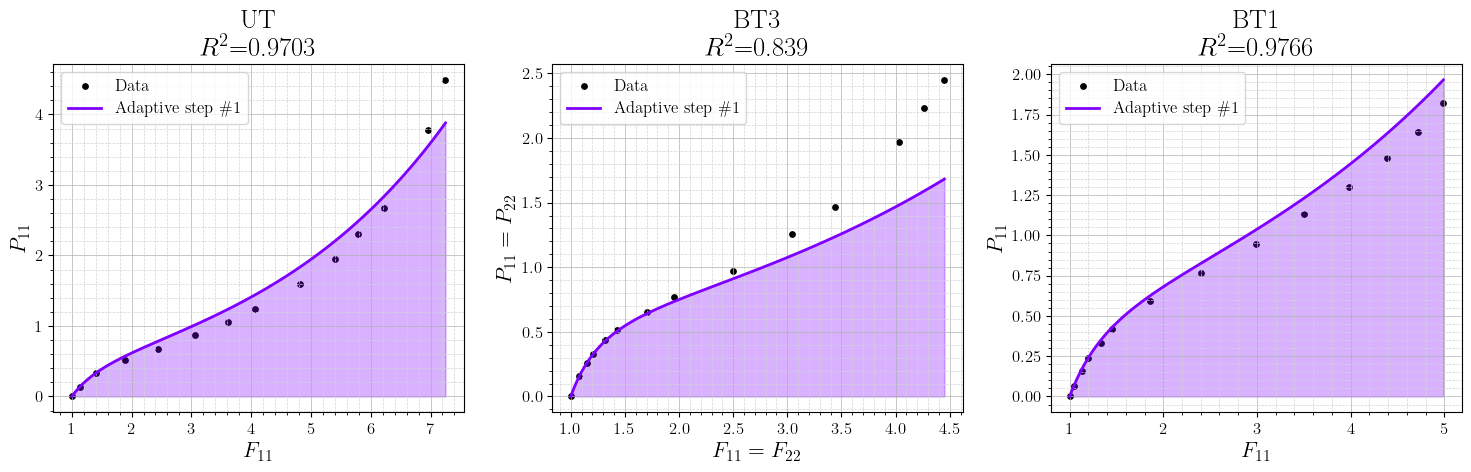}
\caption{
Stress-stretch data and discovered model for rubber at $20^{\circ}\mathrm{C}$ with hyperparameters $n_a = 1$ and $s = 1$, average $R^2=0.9286$.
}
\label{fig:rubber_20_adaptive_1_1}
\end{figure}

To demonstrate the inner workings of the adaptive \MF{} method, we discuss in the following, how increasing the number of features $n_a$ affects the discovery results for the two datasets. Setting $n_a=1$, our method identifies the 1-term model
\begin{equation}
    \tilde W = \num{2.0041e+00} \ \sum_{i=1}^3 \left[ \exp{(0.30 \ [\lambda_i-1])} - 1 \right] \\
\end{equation}
for rubber at $20^{\circ}\mathrm{C}$. As shown in \cref{fig:rubber_20_adaptive_1_1}, the model is in good qualitative agreement with the data. However, the average coefficient of determination $R^2=0.9286$ indicates that there remains potential for improving fitting accuracy. These results are in line with our previous work \citep{martonova_material_2026}, which was restricted to one modeling feature in the strain energy density function.

\begin{figure}[!ht]
\centering
\includegraphics[width=0.75\textwidth]{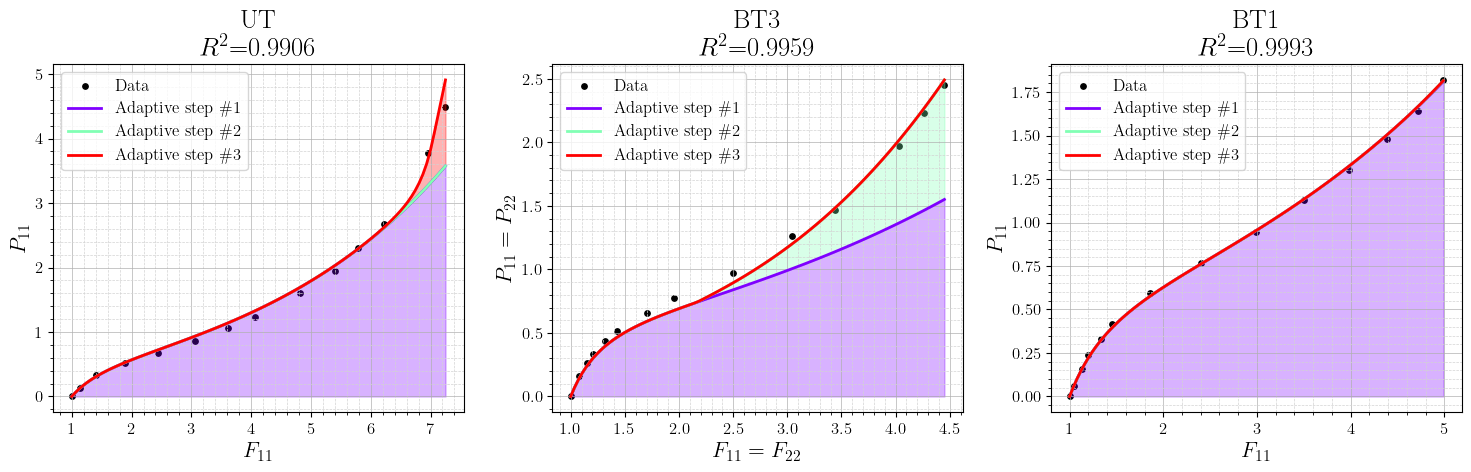}
\caption{
Stress-stretch data and discovered model for rubber at $20^{\circ}\mathrm{C}$ with hyperparameters $n_a = 3$ and $s = 1$, average $R^2= 0.9953 $.
}
\label{fig:rubber_20_adaptive_3_1}
\end{figure}

Next, we increase the number of modeling features in the discovered material model by setting the hyperparameters to $n_a = 3$ and $s = 1$. The adaptive \MF{} method discovers the 3-term model
\begin{equation}
\begin{aligned}
    \tilde W =
    &\num{1.8468e+00} \ \sum_{i=1}^3 \left[ \exp{(0.30 \ [\lambda_i-1])} - 1 \right] \\
    +&\num{2.7204e-12} \ \sum_{j=2}^3 \sum_{k<j} \left[ \exp{(10.58[\lambda_j\lambda_k-1] -22.23)} - \exp{(-22.23)} \right] \\
    +&\num{1.2913e-09} \ \sum_{i=1}^3 \left[ \exp{(5.34[\lambda_i-1] -13.93)} - \exp{(-13.93)} \right],
\end{aligned}
\end{equation}
which yields an average $R^2$ value of $0.9953$. As shown in \cref{fig:rubber_20_adaptive_3_1}, the resulting model response aligns well with the data. In the figure, the contribution of each feature added during the model construction process is indicated by a distinct color. The first modeling feature is equivalent to the previous results with $n_a=1$. The newly added second modeling feature improves the fitting accuracy in equibiaxial tension, while the newly added third feature improves the fitting accuracy in uniaxial tension.

\begin{figure}[!ht]
\centering
\includegraphics[width=0.75\textwidth]{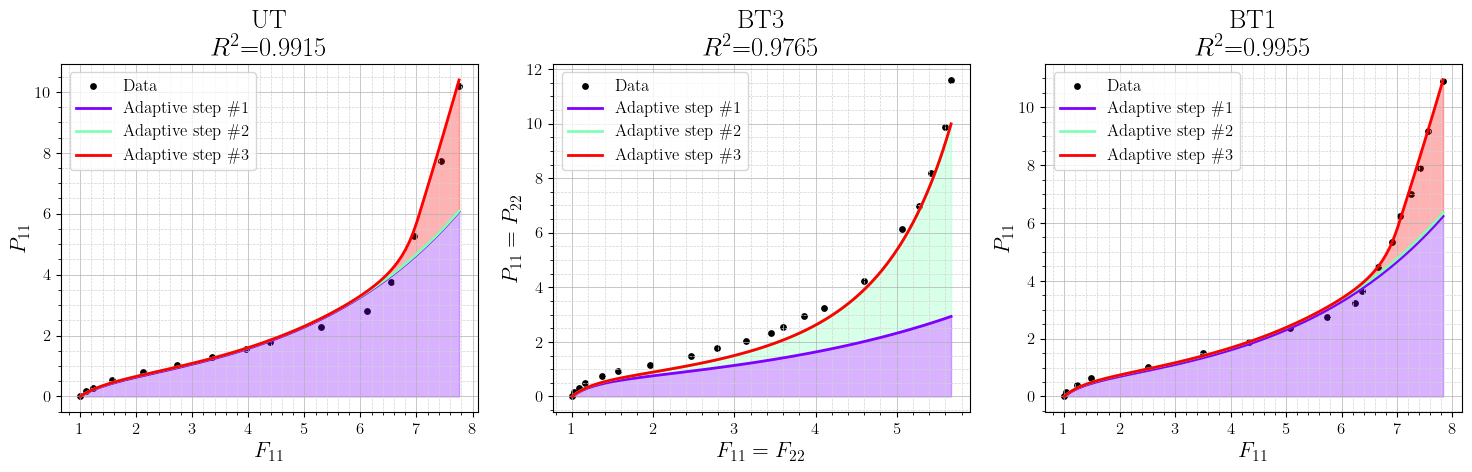}
\caption{
Stress-stretch data and discovered model for rubber at $50^{\circ}\mathrm{C}$ with hyperparameters $n_a = 3$ and $s = 0.7$, average $R^2= 0.9878 $.
}
\label{fig:rubber_50_adaptive_3_0p7}
\end{figure}

The results for rubber at $50^{\circ}\mathrm{C}$ are similar. For the hyperparameters $n_a = 3$ and $s = 0.7$, we discover the 3-term model
\begin{equation}
\begin{aligned}
    \tilde W =
    &\num{1.6411e+00} \ \sum_{i=1}^3 \left[ \exp{(0.35 \ [\lambda_i-1])} - 1 \right] \\
    +&\num{4.9078e-07} \ \sum_{j=2}^3 \sum_{k<j} \left[ \exp{(0.10[\lambda_j\lambda_k-1] +13.93)} - \exp{(13.93)} \right] \\
    +&\num{2.2080e-09} \ \sum_{i=1}^3 \left[ \exp{(4.03[\lambda_i-1] -5.63)} - \exp{(-5.63)} \right],
\end{aligned}
\end{equation}
which yields an average $R^2$ value of $0.9878$. Interestingly, we discover the same modeling features as for rubber at $20^{\circ}\mathrm{C}$, albeit with different material parameters. \Cref{fig:rubber_50_adaptive_3_0p7} shows that the model response aligns well with the data. We observe that the second term is necessary for describing equibiaxial tension, and the third term is necessary for describing uniaxial tension and pure shear.

As described in \cref{sec:polyconvexity}, polyconvexity can be enforced in the discovered material model through a simple configuration choice in the implementation, whereby modeling features that are not guaranteed to satisfy polyconvexity are systematically excluded. For the rubber datasets and the chosen hyperparameters, adaptive \MF{} discovers the same models regardless of whether polyconvexity is enforced. Thus, for the present example, imposing the polyconvexity constraint does not affect the fitting accuracy of the method. We finally note that the fitting accuracy achieved with adaptive \MF{} is nearly the same as using a generalized-invariant-based constitutive neural network for the same data \citep{martonova_generalized_2025}.

\subsection{Anisotropic skin}

\begin{figure}[!ht]
    \centering
    \includegraphics[width=0.4\textwidth]{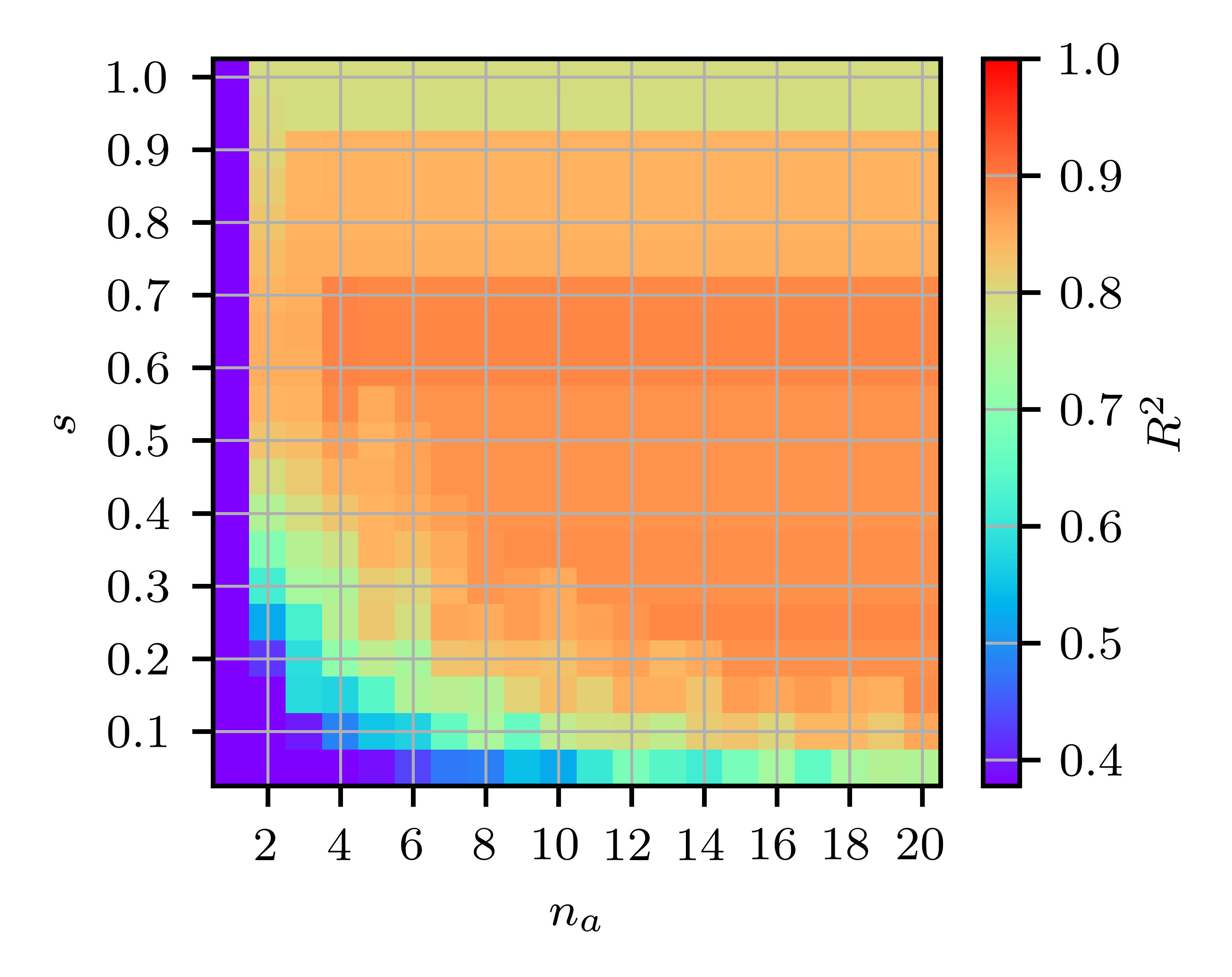}
    \caption{
    Hyperparameter analysis for adaptive \MF{} applied to skin data.
    }
\label{fig:pig_skin_hyperparameter_plot}
\end{figure}

We next apply the adaptive \MF{} framework to experimental data obtained from anisotropic skin. The data are taken from \cite{linka_automated_2023-1} and provided in \cref{app: experimental data}. We employ the anisotropic feature database described in \cref{sec:adaptive fingerprint database} and we again perform a systematic hyperparameter analysis, in which we vary both $n_a$ and $s$. The resulting $R^2$ values are shown in \cref{fig:pig_skin_hyperparameter_plot}. We observe a sweet spot for the hyperparameter value $s = 0.7$, which is kept fixed in the following examples.

\begin{figure}[ht]
    \centering
    \includegraphics[width=\textwidth]{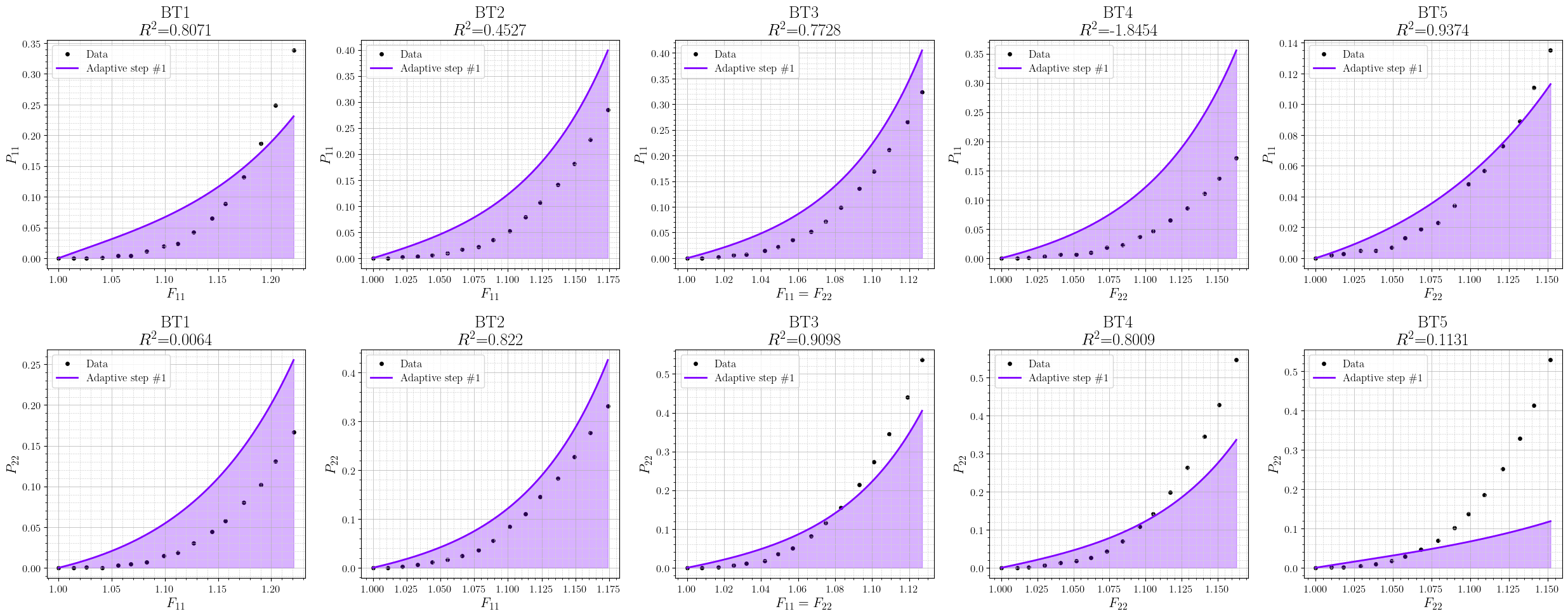}
    \caption{%
        Stress-stretch data and discovered model for skin with hyperparameters $n_a = 1$ and $s = 0.7$, average $R^2=0.3777$.
        }
\label{fig:pig_skin_adaptive_1_0p7}
\end{figure}

Due to its anisotropic characteristics, the skin dataset is particularly well suited for examining the impact of the hyperparameter $n_a$. For $n_a = 1$, the adaptive \MF{} framework identifies an isotropic 1-term model
\begin{equation}
\begin{aligned}
\tilde W = \num{3.4774e-03} \sum_{j=2}^3 \sum_{k<j}  [\exp{(9.08[\lambda_j\lambda_k-1])}], \label{eq:pig_skin_adaptive_1}
\end{aligned}
\end{equation}
which shows poor predictive performance, with an average coefficient of determination of $R^2 = 0.3777$,~see \cref{fig:pig_skin_adaptive_1_0p7}. 

Thus, a purely isotropic formulation cannot represent the pronounced anisotropic response of skin. For $n_a = 2$, the framework adds one anisotropic term and discovers the following transversely isotropic 2-term model
\begin{equation}
\begin{aligned}
    \tilde W =
    +&\num{2.3271e-03} \ \sum_{j=2}^3 \sum_{k<j} \left[ \exp{(9.08[\lambda_j\lambda_k-1])} \right] \\
    +&\num{7.4216e+02} \ \log (\cosh (0.60 \ [\lambda_a-1]))^2. \label{eq:pig_skin_adaptive_2_0p7}
\end{aligned}
\end{equation}

\begin{figure}[ht]
    \centering
    \includegraphics[width=\textwidth]{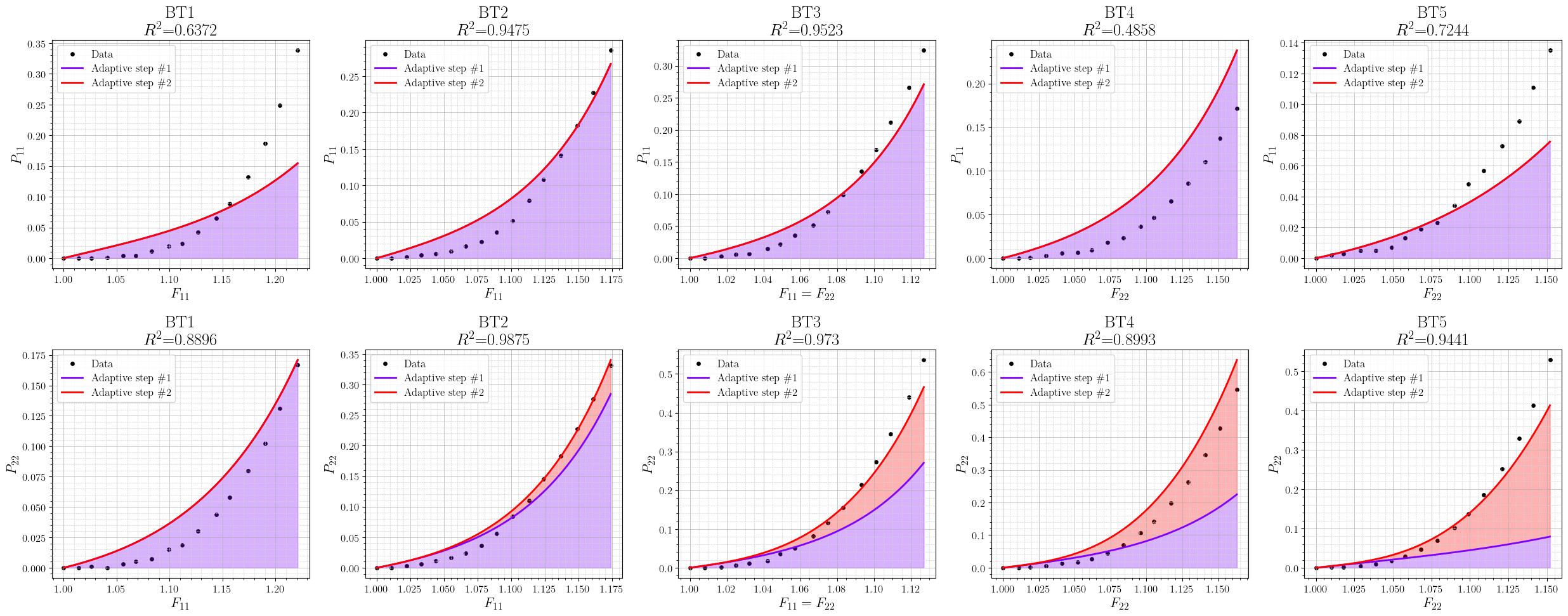}
    \caption{%
        Stress-stretch data and discovered model for skin with hyperparameters $n_a = 2$ and $s = 0.7$, average $R^2=0.8441$.
        }
\label{fig:pig_skin_adaptive_2_0p7}
\end{figure}

This model achieves a substantially improved average accuracy of $R^2 = 0.8441$,~see \cref{fig:pig_skin_adaptive_2_0p7}, and thereby demonstrates that anisotropic features are essential for capturing the mechanical behavior of the skin. 
Choosing $n_a=4$ yields the best hyperparameter combination regarding the balance of model complexity and accuracy. The resulting 4-term model with an average $R^2$ value of $0.8916$ reads as

\begin{equation}
\begin{aligned}
    \tilde W =
    &\num{2.0062e-03} \ \sum_{j=2}^3 \sum_{k<j} \left[ \exp{(9.08 \ [\lambda_j\lambda_k-1])} - 1 \right] \\
    +&\num{4.4787e+02} \ \log (\cosh (0.60 \ [\lambda_a-1]))^2 \\
     +&\num{3.7148e-10} \ \sum_{j=2}^3 \sum_{k<j} \left[ \exp{(25.00[\lambda_j\lambda_k-1] +11.17)} - \exp{(11.17)} \right] \\
    +&\num{1.4694e+03} \ \log (\cosh (0.40 \ [\lambda_a-1]))^2. \label{eq:pig_skin_adaptive_4_0p7}
\end{aligned}
\end{equation}

\begin{figure}[ht]
    \centering
    \includegraphics[width=\textwidth]{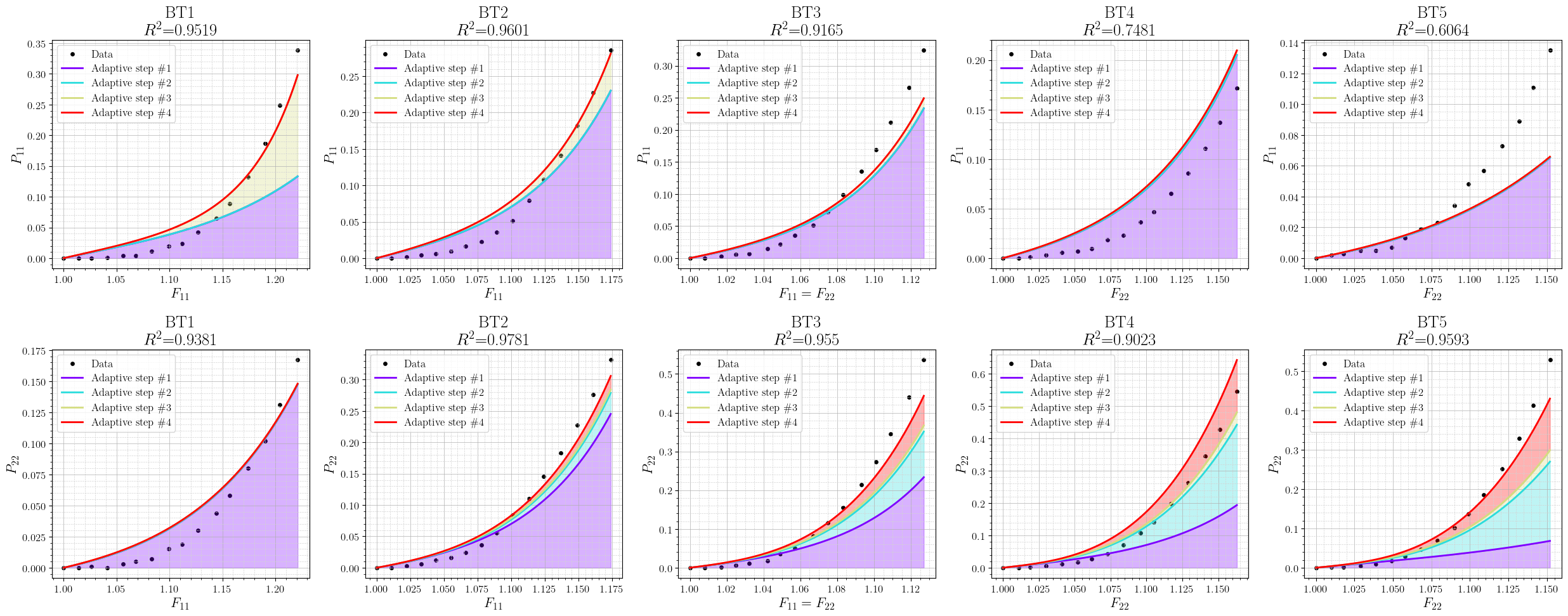}
    \caption{%
        Stress-stretch data and discovered model for skin with hyperparameters $n_a = 4$ and $s = 0.7$, average $R^2=0.8916$.
        }
\label{fig:pig_skin_adaptive_4_0p7}
\end{figure}

\cref{fig:pig_skin_adaptive_4_0p7} illustrates the corresponding mechanical response. Overall, we obtain satisfactory agreement between the experimental observations and the model predictions. However, for certain loading modes and stress components, deviations remain. In particular, the stress component $P_{11}$ of loading mode BT5 is not captured accurately.
Several factors may contribute to these discrepancies. The selected feature functions may not be sufficiently expressive, the assumed class of anisotropy may be overly restrictive, the fiber direction may not be perfectly aligned with the $\bfe_2$-direction in the experiment, or the adaptive algorithm may not have identified the optimal combination of features. Additionally, measurement uncertainties in the experimental data cannot be excluded.
Notably, the proposed adaptive \MF{} approach achieves a fitting accuracy comparable to that of constitutive artificial neural networks, as reported in \cite{linka_automated_2023-1}. Importantly, the neural network models exhibit similar difficulties in simultaneously fitting all loading modes, particularly with respect to the stress component $P_{11}$ of loading mode BT5. This observation suggests that the remaining discrepancies are more likely attributable to modeling assumptions or experimental uncertainties than to insufficient expressiveness of the adaptive \MF{} feature database.

In contrast to the isotropic material considered in the previous section, a key difference emerges when increasing $n_a$. 
While the 1-term model in \cref{eq:pig_skin_adaptive_1} and the corresponding standardized experiments in \cref{fig:pig_skin_adaptive_1_0p7} yield a purely isotropic response, increasing $n_a$ introduces anisotropic contributions to the first Piola-Kirchhoff stress tensor through the stretch $\lambda_a$.
To understand the effect of this contribution, we evaluate the anisotropic part of the stress,
\begin{equation}
    \tilde{\bfP}^{\mathrm{ani}} = \frac{\partial\tilde{W}}{\partial\lambda_a} \frac{\partial\lambda_a}{\partial\bfF} 
    = \frac{\partial\tilde{W}}{\partial\lambda_a}\, \begin{pmatrix}
        0 & 0 & 0 \\
        0 & 1 & 0 \\
        0 & 0 & 0
    \end{pmatrix},
\end{equation}
where we used that the stretch in fiber direction $\lambda_a$ coincides with $F_{22}$. A detailed derivation of the stress tensor is provided in \cref{sec:theory}.
This expression shows that anisotropic contributions act exclusively on the $P_{22}$ component of the first Piola-Kirchhoff stress tensor. Consequently, the additional terms introduced in the second adaptive step in \cref{eq:pig_skin_adaptive_2_0p7}, as well as the second and fourth terms in \cref{eq:pig_skin_adaptive_4_0p7}, affect only this stress component. This behavior is consistent with the observed responses in the standardized experiments shown in \cref{fig:pig_skin_adaptive_2_0p7,fig:pig_skin_adaptive_4_0p7}.
It should be noted, however, that this property is specific to the considered standardized deformation modes. In more general settings, e.g., when the deformation gradient contains off-diagonal components or when the fiber direction is not aligned with a principal axis, the anisotropic contribution would, in general, influence all stress components.

Through our systematic increase in the number of model terms, we demonstrate that the choice of the hyperparameter $n_a$ plays a role similar to that of a regularization parameter in continuous optimization--based model discovery approaches \citep{martonova_automated_2024,martonova_discovering_2025a,vervenne_2025,flaschel_non-smooth_2025}. Small values of $n_a$ promote simple models with limited complexity but reduce predictive accuracy. Larger values increase expressiveness but may introduce overfitting. In \cite{martonova_automated_2024}, an increase in the regularization parameter also reduced model complexity and goodness of fit. In the extreme case, the identified model for orthotropic cardiac tissue reduced to an isotropic formulation. This behavior resembles the transverse isotropic skin considered here, where an excessively restrictive choice of $n_a$ likewise suppresses anisotropic contributions and leads to an isotropic model with limited predictive capability. These results show that the parameter $n_a$ provides a transparent and physically interpretable mechanism to balance model complexity and predictive performance within the adaptive \MF{} framework.

Without imposing polyconvexity, the material model discovered by the adaptive \MF{} in \cref{eq:pig_skin_adaptive_2_0p7,eq:pig_skin_adaptive_4_0p7} includes the feature function $h^A$. The discovered model may be polyconvex, but it is not guaranteed to be polyconvex. If we impose polyconvexity a priori, the feature function $h^A$ is not considered, and the resulting discovered material model is polyconvex. However, as a result of the polyconvexity constraint, the fitting accuracy for $n_a=4$ and $s=0.7$ reduces from an average coefficient of determination of $0.8916$ to $0.8371$.

\section{Conclusions}
\label{sec:conclusions}
In this work, we have introduced an adaptive extension of the Material Fingerprinting framework for the rapid discovery of anisotropic constitutive models in incompressible hyperelasticity. Building on the original lookup table-based concept, the proposed strategy enables the discovery not only of predefined material models contained in a fixed database, but also of arbitrary linear combinations of physically admissible feature functions. By coupling an adaptive feature database with an iterative pattern recognition algorithm, we obtain an accurate yet computationally efficient methodology for constitutive model discovery.

The adaptive formulation preserves the key advantages of Material Fingerprinting: the separation into an offline database generation stage and an online identification stage avoids the solution of continuous, potentially non-convex optimization problems. In the online phase, model discovery reduces to a sequence of algebraic operations and similarity searches within a precomputed database, which ensures robustness and real-time capability. At the same time, the adaptive enrichment strategy significantly increases modeling flexibility and accuracy. By successively adding features to the strain energy density function, the user can fine-tune the balance between model sparsity and model accuracy without having to rely on regularization. As such, the framework recovers complex multi-term formulations, including structures equivalent to multi-term Ogden-type and Holzapfel-Gasser-Ogden-type models.

From a theoretical perspective, the selected feature functions are formulated in terms of invariant quantities and are consistent with the fundamental principles of objectivity and material symmetry. Enforcing polyconvexity is straightforward by restricting the admissible feature set in the online phase.

The application to experimental datasets of isotropic rubber and anisotropic skin demonstrates the performance and versatility of the approach. For rubber, adaptive Material Fingerprinting can substantially improve the coefficient of determination compared to the original method by successively adding more terms to the model and achieve accuracies comparable to constitutive artificial neural networks without requiring a costly training phase. For skin, the framework successfully identifies the need to include anisotropic features and recovers model structures consistent with established formulations. A hyperparameter study highlights that the user can modulate model complexity and successively include additional features to build the final model bottom-up through iteration across possible fingerprints. This is an advantage over most optimization-based approaches that successively remove irrelevant features and build the model top-down through regularization terms in the optimization function.

Overall, adaptive Material Fingerprinting combines interpretability, physical consistency, and computational efficiency in a unified framework for automated model discovery. It offers a scalable alternative to optimization-based and purely data-driven approaches, particularly in scenarios where rapid and robust material characterization is required. Future work may extend the database to additional deformation modes, explore inelastic material behavior, incorporate alternative feature sets, and further investigate automated strategies for hyperparameter tuning and model selection.

\section*{Code and data availability}

The code and data are available at
\begin{center}
    \url{https://github.com/Material-Fingerprinting/material-fingerprinting-adaptive} .
\end{center}
% The isotropic adaptive database is added to the pip-installable Python package for Material Fingerprinting \citep{flaschel_python_2025}, which is published on GitHub under
% \begin{center}
%     \url{https://github.com/Material-Fingerprinting/material-fingerprinting} .
% \end{center}

\section*{Acknowledgments}

The authors acknowledge support from the European Research Council (ERC) Grant 101141626 DISCOVER. Funded by the European Union. Views and opinions expressed are, however, those of the authors only and do not necessarily reflect those of the European Union or the European Research Council
Executive Agency. Neither the European Union nor the granting authority can be held responsible for them.
% The authors acknowledge the scientific support and HPC resources provided by the Erlangen National High Performance Computing Center NHR@FAU of the Friedrich-Alexander-Universität Erlangen-Nürnberg (FAU). The hardware is funded by the German Research Foundation (DFG).
% The authors utilized generative artificial intelligence to enhance the writing style in certain sections of the manuscript.
% After using these tools, the authors reviewed and edited the content as needed and take full responsibility for the content of the publication.

% \clearpage

\appendix
\crefalias{section}{appendix}

% \section{Compressible material behavior}

% \begin{equation}
%     W(\bfF) =
%     \sum_{l=1}^{n_l} \left[
%     \theta^{\text{I}}_l \sum_{i=1}^3 f_l(\lambda_i;\bfalpha^{\text{I}}_l)
%     + \theta^{\text{II}}_l \sum_{i=1,\ i \neq j}^3\sum_{j=1}^3 f_l(A_{ij};\bfalpha^{\text{II}}_l)
%     + \theta^{\text{III}}_l  f_l(V;\bfalpha^{\text{III}}_l)
%     \right] + \sum_{i=1}^3 f_\mathrm{gr}(\lambda_i),
%     \label{eq:W_energy}
% \end{equation}

\section{Theoretical remarks}
\label{sec:theory}

We develop an adaptive framework that increases the flexibility of the original \MF{} method by introducing a generic set of feature functions defining the strain energy density. In this appendix, we address physical admissibility constraints, the derivation of the first Piola-Kirchhoff stress tensor, and the relation to polyconvexity. Central to this discussion is the additive decomposition of the isochoric strain energy in \cref{eq:W_energy_incompressible} into isotropic and anisotropic contributions
\begin{equation}
    \tilde{W}(\bfF, \bfa; \bftheta, \bfalpha) = \tilde{W}^{\rm iso}(\bfF; \bftheta, \bfalpha) + \tilde{W}^{\rm ani}(\bfF, \bfa; \bftheta, \bfalpha).
    \label{eq:W_decompose}
\end{equation}

\subsection{Physical admissibility}
\label{sec:physical_admissibility}

Irrespective of the particular choice of the strain energy function and, in the present work, the selected set of feature functions, any constitutive formulation must satisfy the material principles of objectivity and material symmetry. To account for anisotropic material behavior, we follow the approach of \cite{boehler_irreducible_1977,boehler_simple_1979} and extend the argument list of the strain energy function by introducing a structural direction $\bfa$ in addition to the deformation gradient $\bfF$.

Objectivity requires invariance of the strain energy density under superposed rigid body rotations and is expressed as
\begin{equation}
    W(\bfQ \bfF, \bfa) = W(\bfF, \bfa),
    \qquad
    \bfP(\bfQ \bfF, \bfa) = \bfQ\,\bfP(\bfF, \bfa),
    \qquad
    \bfQ \in \mathrm{SO}(3),
\end{equation}
where $\bfP$ denotes the first Piola-Kirchhoff stress tensor and $\mathrm{SO}(3)$ the special orthogonal group.
Material symmetry expresses invariance with respect to transformations belonging to the symmetry group of the material. For the extended argument list $(\bfF,\bfa)$, this invariance is required for all rotations $\bfQ\in\mathrm{SO}(3)$ that leave the structural direction invariant, that is, for all $\bfQ$ satisfying $\bfQ^T\bfa=\bfa$. Accordingly, material symmetry takes the form
\begin{equation}
    W(\bfF \bfQ, \bfQ^T\bfa) = W(\bfF,\bfa),
    \qquad
    \bfP(\bfF \bfQ, \bfQ^T\bfa) = \bfP(\bfF,\bfa)\,\bfQ.
\end{equation}
In combination, the principles of objectivity and material symmetry typically imply that the strain energy density can be represented as a scalar-valued isotropic function.

We now consider how these principles are satisfied by the strain energy density $\tilde{W}$. 
The isotropic feature functions are expressed in terms of the principal stretches, which are obtained from the singular value decomposition of the deformation gradient,
\begin{equation}
    \bfF = \bfPhi\, \bfLambda\, \bfOmega^T,
    \label{eq:Singular_F}
\end{equation}
where $\bfPhi$ and $\bfOmega$ are orthogonal tensors and $\bfLambda = \mathrm{diag}(\lambda_1, \lambda_2, \lambda_3)$ contains the strictly positive principal stretches.
Left multiplication of $\bfF$ by an orthogonal tensor affects only $\bfPhi$, while the principal stretches and $\bfOmega$ remain unchanged. 
In an analogous manner, right multiplication modifies only $\bfOmega$. 
As a consequence, the isotropic contribution to the strain energy density in \cref{eq:W_decompose} depends exclusively on the principal stretches and therefore satisfies both objectivity and material symmetry.

For the anisotropic contribution, the feature functions depend solely on the scalar quantities $\lambda_{\bfa}>0$ and $A_{\bfa}>0$. 
Evaluating their transformation behavior under superposed rigid body rotations and material symmetry transformations yields
\begin{equation}
    \lambda_{\bfa}(\bfQ\bfF,\bfa) = \lambda_{\bfa}(\bfF,\bfa), \qquad
    A_{\bfa}(\bfQ\bfF,\bfa) = A_{\bfa}(\bfF,\bfa),
\end{equation}
and
\begin{equation}
    \lambda_{\bfa}(\bfF\bfQ,\bfQ^T\bfa) = \lambda_{\bfa}(\bfF,\bfa), \qquad
    A_{\bfa}(\bfF\bfQ,\bfQ^T\bfa) = A_{\bfa}(\bfF,\bfa).
\end{equation}
These relations demonstrate that the anisotropic part of the strain energy density also satisfies the requirements of objectivity and material symmetry.

\subsection{Stress calculation} \label{sec:app_stress_calculation}

\CHANGE{In direct analogy to the decomposition of the isochoric strain energy $\tilde W$ in \cref{eq:W_decompose}, the stress contribution corresponding to $\tilde W$ can be written as the sum of an isotropic part and an anisotropic part}
\begin{equation}
    \CHANGE{\tilde{\bfP} = \frac{\partial \tilde W}{\partial \bfF} = \frac{\partial \tilde W^{\rm iso}}{\partial \bfF} + \frac{\partial \tilde W^{\rm ani}}{\partial \bfF} = \tilde{\bfP}^{\rm iso} + \tilde{\bfP}^{\rm ani}.}
\end{equation}
We first consider the isotropic contribution. Making use of the singular value representation of the deformation gradient and of \cref{eq:stress,eq:W_decompose}, the isotropic stress tensor assumes the form
\begin{equation}
    \tilde{\bfP}^{\rm iso}
    =
    \bfPhi\,
    \mathrm{diag}\!\left(
        \frac{\partial \tilde{W}^{\rm iso}}{\partial \lambda_1},
        \frac{\partial \tilde{W}^{\rm iso}}{\partial \lambda_2},
        \frac{\partial \tilde{W}^{\rm iso}}{\partial \lambda_3}
    \right)
    \bfOmega^T.
    \label{eq:stress_iso}
\end{equation}
Consequently, the computation of $\tilde{\bfP}^{\rm iso}$ reduces to evaluating the derivatives of the isotropic feature functions $\tilde{w}^{\text{I}}$ and $\tilde{w}^{\text{II}}$ introduced in \cref{eq:w12}. For the derivative with respect to a principal stretch $\lambda_j$ one obtains
\begin{equation}
    \frac{\partial\tilde{w}^{\text{I}}(\lambda_1,\lambda_2,\lambda_3;\bfalpha)}{\partial\lambda_j}
    =
    \left.\frac{\partial g(x;\bfalpha)}{\partial x}\right|_{x=\lambda_j},
\end{equation}
and
\begin{equation}
    \frac{\partial\tilde{w}^{\text{II}}(\lambda_1,\lambda_2,\lambda_3;\bfalpha)}{\partial\lambda_j}
    =
    \sum_{\substack{k=1 \\ k\neq j}}^{3}
    \left.\frac{\partial g(x;\bfalpha)}{\partial x}\right|_{x=\lambda_j\lambda_k}\,\lambda_k.
\end{equation}
As already discussed for the strain energy density, the representation in \cref{eq:stress_iso} immediately implies that the isotropic stress contribution satisfies objectivity and material symmetry. 
Left multiplication of $\bfF$ by a rotation only modifies $\bfPhi$, while right multiplication affects $\bfOmega$ without changing the diagonal term and vice versa for right multiplication, such that $\tilde{\bfP}^{\rm iso}(\bfQ\bfF)=\bfQ\,\tilde{\bfP}^{\rm iso}(\bfF)$ and $\tilde{\bfP}^{\rm iso}(\bfF\bfQ)=\tilde{\bfP}^{\rm iso}(\bfF)\,\bfQ$.

We now turn to the anisotropic contribution. Since the anisotropic strain energy depends solely on the scalar quantities $\lambda_{\bfa}$ and $A_{\bfa}$, the corresponding stress contribution follows from the chain rule as
\begin{equation}
    \tilde{\bfP}^{\rm ani}
    =
    \frac{\partial \tilde{W}^{\rm ani}}{\partial \lambda_{\bfa}}
    \frac{\partial \lambda_{\bfa}}{\partial \bfF}
    +
    \frac{\partial \tilde{W}^{\rm ani}}{\partial A_{\bfa}}
    \frac{\partial A_{\bfa}}{\partial \bfF}.
\end{equation}
Using the definitions of the anisotropic feature functions in \cref{eq:w34}, the required derivatives are given by
\begin{equation}
    \frac{\partial\tilde{w}^{\text{III}}}{\partial\lambda_{\bfa}}
    =
    \frac{\partial h(\lambda_{\bfa};\bfalpha)}{\partial\lambda_{\bfa}},
    \qquad
    \frac{\partial \lambda_{\bfa}}{\partial\bfF}
    =
    \frac{1}{\lambda_{\bfa}}\,[\bfF\bfa]\otimes\bfa,
\end{equation}
and
\begin{equation}
    \frac{\partial\tilde{w}^{\text{IV}}}{\partial A_{\bfa}}
    =
    \frac{\partial h(A_{\bfa};\bfalpha)}{\partial A_{\bfa}},
    \qquad
    \frac{\partial A_{\bfa}}{\partial\bfF}
    =
    \left[
        A_{\bfa}\,\bfI
        -
        \frac{1}{A_{\bfa}}\,
        [\mathrm{cof}(\bfF)\,\bfa]\otimes[\mathrm{cof}(\bfF)\,\bfa]
    \right]\bfF^{-T}.
\end{equation}
From the transformation properties of $\lambda_{\bfa}$ and $A_{\bfa}$ with respect to rotations of the deformation gradient and the fiber direction, it follows that the anisotropic stress contribution satisfies material symmetry and objectivity in the sense that $\tilde{\bfP}^{\rm ani}(\bfQ\bfF,\bfa) = \bfQ\,\tilde{\bfP}^{\rm ani}(\bfF,\bfa)$ and $\tilde{\bfP}^{\rm ani}(\bfF\bfQ,\bfQ^T\bfa) = \tilde{\bfP}^{\rm ani}(\bfF,\bfa)\,\bfQ$.
\begin{equation}
    \tilde{\bfP}^{\rm ani}(\bfQ\bfF,\bfa)
    =
    \bfQ\,\tilde{\bfP}^{\rm ani}(\bfF,\bfa)
    \quad \text{and} \quad
    \tilde{\bfP}^{\rm ani}(\bfF\bfQ,\bfQ^T\bfa)
    =
    \tilde{\bfP}^{\rm ani}(\bfF,\bfa)\,\bfQ.
\end{equation}
Thus, both the isotropic and anisotropic parts of the constitutive response are fully consistent with the underlying symmetry requirements.
We finally note that the Lagrange multiplier $p$ is derived from the plane stress constraint $P_{33}=0$ to compute the components of $\bfP$ for all experiments considered in this work.

\subsection{Polyconvexity}
\label{sec:polyconvexity}

Next, we comment on the role of polyconvexity within the proposed adaptive \MF{} framework. 
Polyconvexity, originally introduced by \cite{ball_convexity_1976}, together with coercivity, provides a sufficient mathematical framework for the existence of minimizers of elastic energy functionals at finite strains, see \cite{ciarlet_mathematical_1988}. 
However, in contrast to objectivity and material symmetry, polyconvexity does not constitute a physical principle and therefore does not represent a mandatory requirement for constitutive laws. 
In the following, we briefly recall the concept of polyconvexity and discuss how it can be readily enforced in the adaptive \MF{} approach.

\CHANGE{In continuum mechanics, a wide range of anisotropic materials are described not only by the deformation gradient $\bfF$, but also by additional structural tensors.
For the special case of transverse isotropy, this is commonly achieved by introducing a first-order structural tensor $\bfa$, aligned with the preferred material direction orthogonal to the plane of isotropy.
Within this framework, the corresponding strain energy function $W$ is called polyconvex if it admits a representation of the form}
\begin{equation}
    W(\bfF,\bfa)
    =
    \bar{W}(\bfF,\mathrm{cof}(\bfF),\det(\bfF);\bfa),
\end{equation}
where $\bar{W}$ is convex with respect to its arguments $\bfF$, $\mathrm{cof}(\bfF)$, and $\det(\bfF)$.
\CHANGE{Here, $\bar{W}$ captures the general case of compressible materials.
For incompressible materials, the deformation is constrained by $\det(\bfF)=1$, and the strain energy is defined on this restricted set. Accordingly, the polyconvexity condition is formulated without dependence on $\det(\bfF)$, see \cite{ball1977constitutive}.}
The fiber direction $\bfa$ is separated by a semicolon to emphasize that convexity with respect to $\bfa$ is not required.

We first consider the isotropic part of the strain energy. 
\CHANGE{Based on the construction principle introduced in \cref{eq:strain_energy_density_linear_combination} for the adaptive \MF{}, the strain energy is expressed as a linear combination of individual contributions.}
Since \CHANGE{these contributions are entirely expressed in terms of} singular values\CHANGE{, cf. \cref{eq:w12}}, we recall a standard result from matrix convex analysis stating that a function of the form
\begin{equation}
    y^{\rm iso}(\bfY)
    =
    \sum_{i=1}^{p} \chi\bigl(s_i(\bfY)\bigr)
\end{equation}
is convex in $\bfY\in\mathbb{R}^{m\times n}$ if $\chi$ is convex and monotonically non-decreasing, where $p=\min(m,n)$ and $s_i(\bfY)$ denote the singular values of $\bfY$, see \cite{horn_matrix_2012}. 
Noting that the principal stretches $\lambda_i$ are the singular values of $\bfF$ and that the products $\lambda_j\lambda_k$ correspond to the singular values of $\mathrm{cof}(\bfF)$, the isotropic feature functions in \cref{eq:w12} can be written as
\begin{equation}
    \tilde{w}^{\text{I}} =
    \sum_{i=1}^{3} g\bigl(s_i(\bfF);\bfalpha\bigr), \quad
    \tilde{w}^{\text{II}} =
    \sum_{i=1}^{3} g\bigl(s_i(\mathrm{cof}(\bfF));\bfalpha\bigr).
\end{equation}
% \begin{equation}
%     \tilde{w}^{\text{I}} + \tilde{w}^{\text{II}}
%     =
%     \sum_{i=1}^{3} g\bigl(s_i(\bfF);\bfalpha\bigr)
%     +
%     \sum_{i=1}^{3} g\bigl(s_i(\mathrm{cof}(\bfF));\bfalpha\bigr).
% \end{equation}
The additional scaling by the non-negative homogeneity parameters $\bftheta$ preserves convexity, and since sums of convex functions remain convex, the isotropic contribution to the strain energy is polyconvex provided that $g$ is convex and non-decreasing.
\CHANGE{The above derivation holds only under the assumption of a linearly separable strain energy, as adopted in our adaptive approach. 
For the more general case, where the individual contributions may simultaneously depend on the singular values of $\bfF$ and $\mathrm{cof}(\bfF)$, we refer the interested reader to Theorem 5.2 in \cite{ball_convexity_1976}.}

We now turn to the anisotropic part.
To assess its polyconvexity, we examine the convexity properties of the scalar quantities $\lambda_{\bfa}$ and $A_{\bfa}$. 
To this end, we consider the generic function
\begin{equation}
    y^{\rm ani}(\bfY;\bfa)
    =
    \sqrt{\bfa^T \bfY^T \bfY \bfa}.
\end{equation}
Its second directional derivative in direction $\bfH$ is given by
\begin{equation}
    \frac{\mathrm{d}^2}{\mathrm{d}\epsilon^2}
    \left. y^{\rm ani}(\bfY + \epsilon \bfH;\bfa) \right|_{\epsilon=0}
    =
    \frac{
        [\bfa^T\bfH^T\bfH\bfa][\bfa^T\bfY^T\bfY\bfa]
        -
        [\bfa^T\bfY^T\bfH\bfa]^2
    }{
        [\bfa^T\bfY^T\bfY\bfa]^{3/2}
    }
    \ge 0,
\end{equation}
where the denominator is strictly positive for strictly positive singular values of $\bfY$, and the numerator is non-negative by the Cauchy-Schwarz inequality. 
This shows that $\lambda_{\bfa}$ is convex with respect to $\bfF$, while, by analogous arguments, $A_{\bfa}$ is convex with respect to $\mathrm{cof}(\bfF)$.
Finally, the anisotropic feature functions introduced in \cref{eq:w34} take the form
\begin{equation}
    \tilde{w}^{\text{III}} = h(\lambda_{\bfa};\bfalpha), \quad
    \tilde{w}^{\text{IV}} = h(A_{\bfa};\bfalpha).
\end{equation}
% \begin{equation}
%     \tilde{w}^{\text{III}} + \tilde{w}^{\text{IV}}
%     =
%     h(\lambda_{\bfa};\bfalpha)
%     +
%     h(A_{\bfa};\bfalpha).
% \end{equation}
Since the composition of a convex, non-decreasing function with a convex argument remains convex, the anisotropic contribution to the strain energy is polyconvex as well, provided that $h$ satisfies these properties. 
These findings are consistent with established results in the literature, see for example \cite{schroder_invariant_2003} for a detailed discussion of polyconvexity in anisotropic material models.

\section{Taylor series approximation}
\label{sec:taylor}
We note that some of the considered modeling features require the evaluation of the exponential function, which can lead to overflow errors for large deformations. To circumvent this, we approximate the features by a second order Taylor-series around the point at which the slope of the strain energy reaches a predefined threshold.

% Exponential-type feature functions may exhibit numerical overflow due to rapid growth of the exponential term. 
We consider functions of the form
\begin{equation}
    f(x) = \exp(y(x)) - \exp(y(1)),
\end{equation}
where the inner function $y(x)$ depends on the selected feature. 
In particular, we employ 
$y(x)=\alpha[x-1]$ for $g^B$, 
$y(x)=\alpha_1[x-1]+\alpha_2$ for $g^C$, and 
$y(x)=\alpha\,[\max(\{x-1,0\})]^2$ for $h^B$. 
The derivative is
\begin{equation}
    f'(x) = \exp(y(x))\, y'(x).
\end{equation}
To prevent numerical overflow while maintaining smoothness, an adaptive switching strategy based on the derivative magnitude is used. 
We prescribe a threshold $\tau>0$ such that
\begin{equation}
    |f'(x)| \le \tau.
\end{equation}
Using the analytical form of $f'(x)$, this condition is rewritten as an equivalent upper bound on the exponential argument $y(x)$, defining a limiting argument value $y_0$. 
For arguments within this bound, the function is evaluated exactly. 
If the bound is exceeded, the exponential term is replaced by its second-order Taylor expansion around the corresponding point $x_0$, implicitly defined by $|f'(x_0)|=\tau$,
\begin{equation}
    f(x) \approx f(x_0)
    + \exp(y(x_0))\, y'(x_0)\,(x-x_0)
    + \frac{\exp(y(x_0))}{2}
    \Big( y'(x_0)^2 + y''(x_0) \Big)(x-x_0)^2.
\end{equation}
This ensures controlled growth and local $C^2$ consistency at the switching point.

\section{Experimental data}
\label{app: experimental data}

\cref{tab:data_rubber} summarizes the experimental data for vulcanized rubber reported in \cite{treloar_stress-strain_1944}. 
The biaxial test data for skin, taken from \cite{linka_automated_2023-1}, are provided in \cref{tab:data_skin}.

\begin{table}[!ht]
\centering
\caption{First Piola-Kirchhoff stress in MPa vs. stretch data for rubber  \citep{treloar_stress-strain_1944}.}
\label{tab:data_rubber}
% \vspace{0.2cm}
\small
\renewcommand{\arraystretch}{0.9}
\begin{tabular}{|c|c||c|c||c|c||c|c||c|c||c|c|}
\hline
\multicolumn{6}{|c||}{Rubber 20°C} & \multicolumn{6}{c|}{Rubber 50°C} \\ \hline
\multicolumn{2}{|c||}{UT} & 
\multicolumn{2}{c||}{BT3 / ET} & 
\multicolumn{2}{c||}{BT1 / PS} & 
\multicolumn{2}{c||}{UT} & 
\multicolumn{2}{c||}{BT3 / ET} & 
\multicolumn{2}{c|}{BT1 / PS} \\ \hline
$\lambda_{\text{UT}}$ & $P_{11}$ & 
$\lambda_{\text{ET}}$ & $P_{11}$ & 
$\lambda_{\text{PS}}$ & $P_{11}$ & 
$\lambda_{\text{UT}}$ & $P_{11}$ & 
$\lambda_{\text{ET}}$ & $P_{11}$ & 
$\lambda_{\text{PS}}$ & $P_{11}$ \\ \hline
1.00 & 0.000 & 1.00 & 0.000 & 1.00 & 0.000 & 1.00 & 0.000 & 1.00 & 0.000 & 1.00 & 0.000 \\
1.13 & 0.136 & 1.08 & 0.160 & 1.05 & 0.063 & 1.11 & 0.165 & 1.02 & 0.147 & 1.04 & 0.170 \\
1.41 & 0.334 & 1.15 & 0.260 & 1.13 & 0.158 & 1.23 & 0.289 & 1.08 & 0.298 & 1.23 & 0.399 \\
1.89 & 0.519 & 1.21 & 0.331 & 1.20 & 0.238 & 1.57 & 0.537 & 1.16 & 0.478 & 1.48 & 0.627 \\
2.45 & 0.680 & 1.32 & 0.439 & 1.33 & 0.330 & 2.12 & 0.805 & 1.37 & 0.743 & 2.52 & 1.030 \\
3.06 & 0.866 & 1.43 & 0.514 & 1.45 & 0.417 & 2.73 & 1.030 & 1.57 & 0.921 & 3.51 & 1.490 \\
3.62 & 1.060 & 1.70 & 0.657 & 1.86 & 0.594 & 3.36 & 1.300 & 1.96 & 1.170 & 4.33 & 1.900 \\
4.06 & 1.240 & 1.95 & 0.773 & 2.40 & 0.767 & 3.95 & 1.570 & 2.46 & 1.490 & 5.07 & 2.360 \\
4.82 & 1.600 & 2.50 & 0.968 & 2.99 & 0.945 & 4.39 & 1.790 & 2.79 & 1.780 & 5.74 & 2.740 \\
5.41 & 1.950 & 3.04 & 1.260 & 3.50 & 1.130 & 5.29 & 2.290 & 3.14 & 2.040 & 6.24 & 3.220 \\
5.79 & 2.300 & 3.44 & 1.470 & 3.98 & 1.300 & 6.11 & 2.800 & 3.45 & 2.330 & 6.36 & 3.630 \\
6.23 & 2.680 & 4.03 & 1.970 & 4.39 & 1.480 & 6.54 & 3.750 & 3.60 & 2.530 & 6.65 & 4.490 \\
6.96 & 3.780 & 4.26 & 2.230 & 4.72 & 1.640 & 6.95 & 5.270 & 3.86 & 2.960 & 6.91 & 5.340 \\
7.25 & 4.490 & 4.45 & 2.450 & 4.99 & 1.820 & 7.43 & 7.730 & 4.11 & 3.240 & 7.06 & 6.230 \\
--   &  --   & --   &  --   & --   &  --   & 7.76 & 10.200 & 4.60 & 4.240 & 7.26 & 7.000 \\
--   &  --   & --   &  --   & --   &  --   & --   &   --   & 5.06 & 6.150 & 7.42 & 7.890 \\
--   &  --   & --   &  --   & --   &  --   & --   &   --   & 5.28 & 6.990 & 7.56 & 9.180 \\
--   &  --   & --   &  --   & --   &  --   & --   &   --   & 5.42 & 8.180 & 7.83 & 10.900 \\
--   &  --   & --   &  --   & --   &  --   & --   &   --   & 5.59 & 9.870 & --   &   --   \\
--   &  --   & --   &  --   & --   &  --   & --   &   --   & 5.67 & 11.600& --   &   --   \\
\hline
\end{tabular}
\end{table}

\begin{table}[!ht]
\centering
\caption{First Piola-Kirchhoff stress in MPa vs. stretch data for skin under five biaxial loading protocols \citep{linka_automated_2023-1}.}
\label{tab:data_skin}
\small
\renewcommand{\arraystretch}{0.9}
\resizebox{\textwidth}{!}{
\begin{tabular}{|c|c|c||c|c|c||c|c|c||c|c|c||c|c|c|}
\hline
\multicolumn{3}{|c||}{BT1} &
\multicolumn{3}{c||}{BT2} &
\multicolumn{3}{c||}{BT3} &
\multicolumn{3}{c||}{BT4} &
\multicolumn{3}{c|}{BT5} \\
\hline
\multicolumn{3}{|c||}{$\lambda_2=1$} &
\multicolumn{3}{c||}{$\lambda_2=\sqrt{\lambda_1}$} &
\multicolumn{3}{c||}{$\lambda_1=\lambda_2=\lambda$} &
\multicolumn{3}{c||}{$\lambda_1=\sqrt{\lambda_2}$} &
\multicolumn{3}{c|}{$\lambda_1=1$} \\
\hline

$\lambda_1$ & $P_{11}$ & $P_{22}$ &
$\lambda_1$ & $P_{11}$ & $P_{22}$ &
$\lambda$ & $P_{11}$ & $P_{22}$ &
$\lambda_2$ & $P_{11}$ & $P_{22}$ &
$\lambda_2$ & $P_{11}$ & $P_{22}$ \\
\hline

1.00 & 0.000 & 0.000 & 1.00 & 0.000 & 0.000 & 1.00 & 0.000 & 0.000 & 1.00 & 0.000 & 0.000 & 1.00 & 0.000 & 0.000 \\
1.01 & 0.000 & 0.000 & 1.01 & 0.000 & 0.000 & 1.01 & 0.000 & 0.000 & 1.01 & 0.000 & 0.000 & 1.01 & 0.002 & 0.001 \\
1.03 & 0.000 & 0.001 & 1.02 & 0.002 & 0.003 & 1.02 & 0.003 & 0.001 & 1.02 & 0.001 & 0.001 & 1.02 & 0.003 & 0.001 \\
1.04 & 0.001 & 0.000 & 1.03 & 0.004 & 0.006 & 1.03 & 0.006 & 0.006 & 1.03 & 0.003 & 0.006 & 1.03 & 0.005 & 0.005 \\
1.06 & 0.004 & 0.003 & 1.04 & 0.006 & 0.012 & 1.03 & 0.007 & 0.011 & 1.04 & 0.006 & 0.013 & 1.04 & 0.005 & 0.010 \\
1.07 & 0.004 & 0.005 & 1.06 & 0.009 & 0.017 & 1.04 & 0.014 & 0.018 & 1.05 & 0.007 & 0.019 & 1.05 & 0.007 & 0.017 \\
1.08 & 0.011 & 0.007 & 1.07 & 0.017 & 0.024 & 1.05 & 0.022 & 0.035 & 1.06 & 0.010 & 0.026 & 1.06 & 0.013 & 0.028 \\
1.10 & 0.019 & 0.015 & 1.08 & 0.022 & 0.036 & 1.06 & 0.036 & 0.051 & 1.07 & 0.018 & 0.044 & 1.07 & 0.019 & 0.046 \\
1.11 & 0.023 & 0.019 & 1.09 & 0.036 & 0.056 & 1.07 & 0.051 & 0.082 & 1.08 & 0.023 & 0.070 & 1.08 & 0.023 & 0.069 \\
1.13 & 0.043 & 0.030 & 1.10 & 0.052 & 0.084 & 1.08 & 0.072 & 0.116 & 1.10 & 0.036 & 0.108 & 1.09 & 0.034 & 0.101 \\
1.14 & 0.065 & 0.044 & 1.11 & 0.079 & 0.110 & 1.08 & 0.099 & 0.156 & 1.10 & 0.047 & 0.141 & 1.10 & 0.048 & 0.137 \\
1.16 & 0.089 & 0.058 & 1.12 & 0.108 & 0.145 & 1.09 & 0.136 & 0.214 & 1.12 & 0.066 & 0.198 & 1.11 & 0.057 & 0.186 \\
1.17 & 0.132 & 0.080 & 1.14 & 0.141 & 0.183 & 1.10 & 0.169 & 0.273 & 1.13 & 0.085 & 0.263 & 1.12 & 0.073 & 0.252 \\
1.19 & 0.187 & 0.102 & 1.15 & 0.182 & 0.228 & 1.11 & 0.212 & 0.344 & 1.14 & 0.110 & 0.345 & 1.13 & 0.089 & 0.330 \\
1.20 & 0.249 & 0.131 & 1.16 & 0.227 & 0.277 & 1.12 & 0.266 & 0.441 & 1.15 & 0.137 & 0.428 & 1.14 & 0.111 & 0.413 \\
1.22 & 0.338 & 0.167 & 1.17 & 0.285 & 0.332 & 1.13 & 0.324 & 0.536 & 1.16 & 0.171 & 0.547 & 1.15 & 0.135 & 0.529 \\
\hline
\end{tabular}
}
\end{table}

\section{Results for heuristically chosen hyperparameters}
\label{sec:heuristic_hyperparameters}

\begin{figure}[!ht]
\centering
\includegraphics[width=0.75\textwidth]{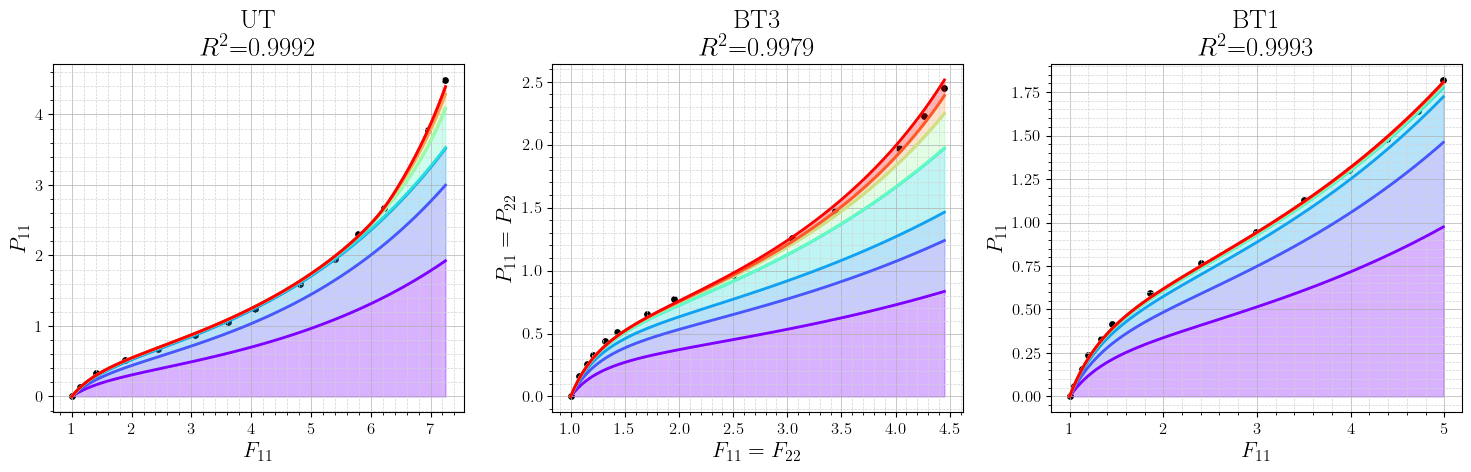}
\caption{
Stress-stretch data and discovered model for rubber at $20^{\circ}\mathrm{C}$ with hyperparameters $n_a = 10$ and $s = 0.5$, average $R^2=0.9988$.
}
\label{fig:rubber_20_adaptive_10_0p5}
\end{figure}

\begin{figure}[!ht]
\centering
\includegraphics[width=0.75\textwidth]{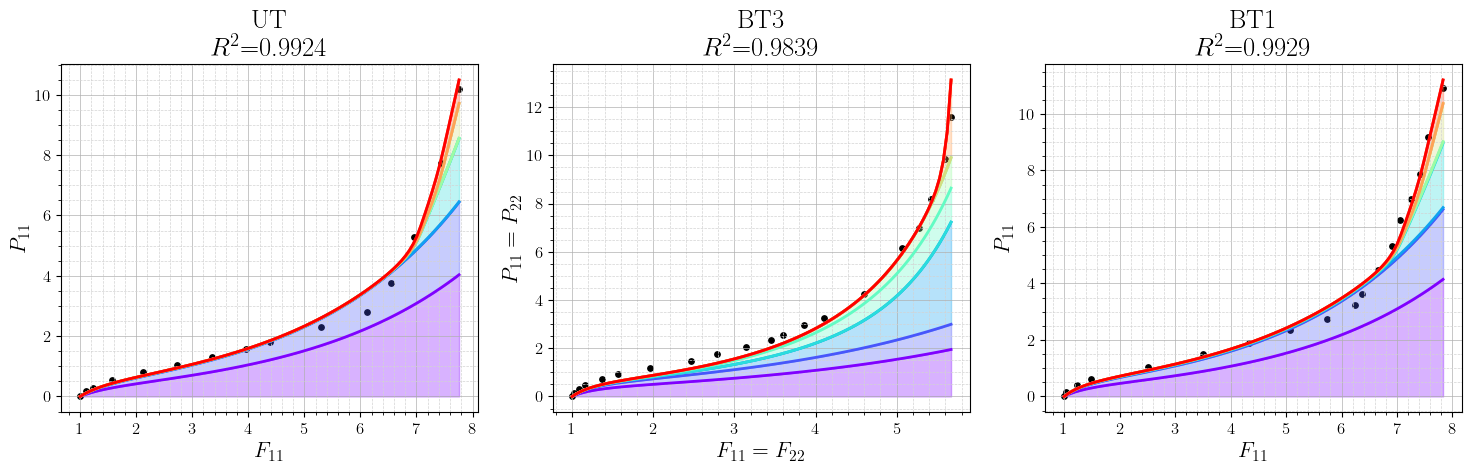}
\caption{
Stress-stretch data and discovered model for rubber at $50^{\circ}\mathrm{C}$ with hyperparameters $n_a = 10$ and $s = 0.5$, average $R^2=0.9897$.
}
\label{fig:rubber_50_adaptive_10_0p5}
\end{figure}

\begin{figure}[ht]
    \centering
    \includegraphics[width=\textwidth]{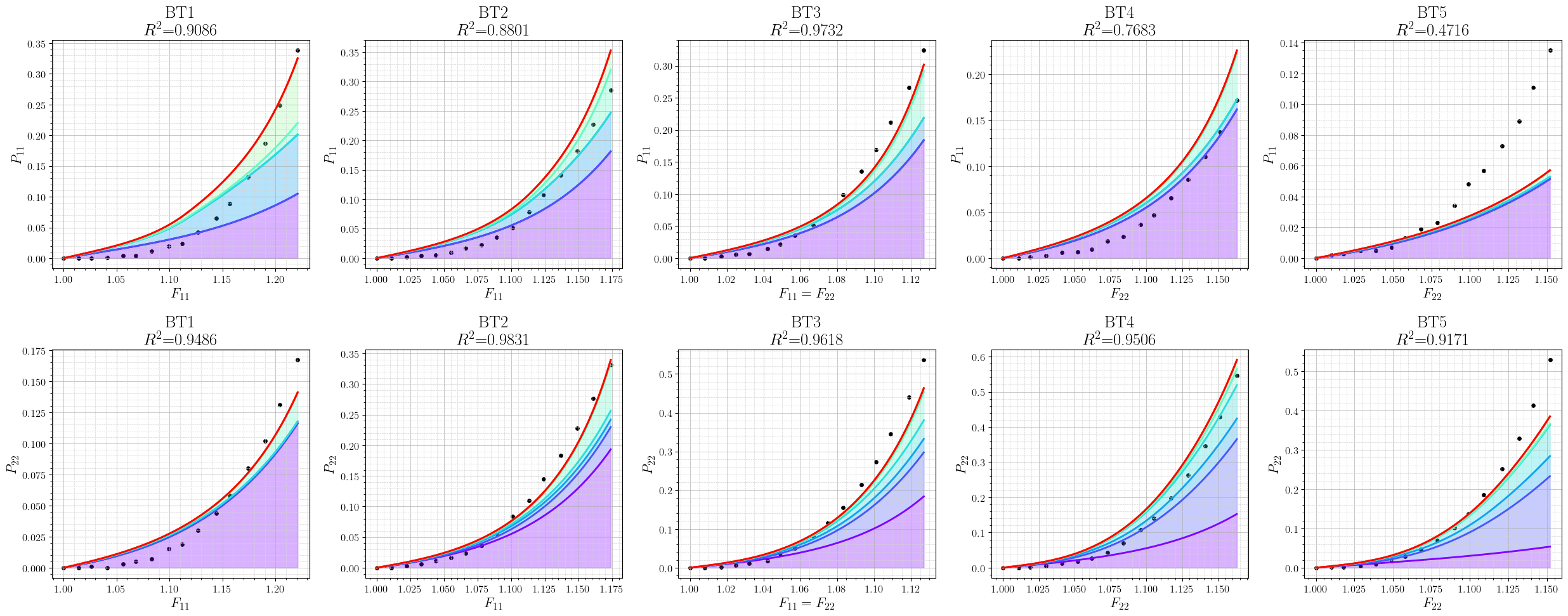}
    \caption{%
        Stress-stretch data and discovered model for skin with hyperparameters $n_a = 10$ and $s = 0.5$, average $R^2=0.8763$.
        }
\label{fig:pig_skin_adaptive_10_0p5}
\end{figure}

As demonstrated in \cref{sec:results}, the hyperparameters of the adaptive \MF{} approach can be tuned to achieve an effective trade-off between predictive accuracy and model complexity. However, when fitting accuracy is the primary objective, our results indicate that the heuristic choice of $n_a=10$ and $s=0.5$ consistently provides highly accurate solutions. \Cref{fig:rubber_20_adaptive_10_0p5,fig:rubber_50_adaptive_10_0p5} show the corresponding results for rubber at $20^{\circ}\mathrm{C}$ and $50^{\circ}\mathrm{C}$, respectively, and \cref{fig:pig_skin_adaptive_10_0p5} shows the corresponding results for the porcine skin data.

%\nocite{*}
\bibliographystyle{elsarticle-harv}
\bibliography{bib_Moritz,bib_Hagen,bib_Denisa}

@article{martonova_discovering_2025a,
  title = {Discovering Dispersion: {{How}} Robust Is Automated Model Discovery for Human Myocardial Tissue?},
  shorttitle = {Discovering Dispersion},
  author = {Martonov{\'a}, Denisa and Leyendecker, Sigrid and Holzapfel, Gerhard A. and Kuhl, Ellen},
  year = 2025,
  month = aug,
  journal = {Biomechanics and Modeling in Mechanobiology},
  issn = {1617-7940},
  doi = {10.1007/s10237-025-02005-x},
  urldate = {2025-08-28},
  langid = {english}
}

@article{vervenne_2025,
	title = {Constitutive Neural Networks for Main Pulmonary Arteries: Discovering the Undiscovered},
	shorttitle = {Constitutive Neural Networks for Main Pulmonary Arteries},
	author = {Vervenne, Thibault and Peirlinck, Mathias and Famaey, Nele and Kuhl, Ellen},
	year = 2025,
	month = feb,
	journal = {Biomechanics and Modeling in Mechanobiology},
	issn = {1617-7940},
	doi = {10.1007/s10237-025-01930-1},
	urldate = {2025-02-26},
	langid = {english}
}

@article{holthusen2026generalized,
title = {A generalized dual potential for inelastic Constitutive Artificial Neural Networks: A JAX implementation at finite strains},
journal = {Journal of the Mechanics and Physics of Solids},
volume = {206},
pages = {106337},
year = {2026},
issn = {0022-5096},
doi = {https://doi.org/10.1016/j.jmps.2025.106337},
url = {https://www.sciencedirect.com/science/article/pii/S0022509625003084},
author = {Hagen Holthusen and Kevin Linka and Ellen Kuhl and Tim Brepols}
}

@article{holthusen2026complement,
title = {A complement to neural networks for anisotropic inelasticity at finite strains},
journal = {Computer Methods in Applied Mechanics and Engineering},
volume = {450},
pages = {118612},
year = {2026},
issn = {0045-7825},
doi = {https://doi.org/10.1016/j.cma.2025.118612},
url = {https://www.sciencedirect.com/science/article/pii/S0045782525008849},
author = {Hagen Holthusen and Ellen Kuhl}
}

@inproceedings{ball1977constitutive,
  title={Constitutive inequalities and existence theorems in nonlinear elastostatics},
  author={Ball, John M},
  booktitle={Nonlinear analysis and mechanics: Heriot-Watt symposium},
  volume={1},
  number={4},
  pages={187--241},
  year={1977},
  organization={Pitman London}
}

@article{efron_least_2004,
	title = {{LEAST} {ANGLE} {REGRESSION}},
	volume = {32},
	language = {en},
	number = {2},
	journal = {The Annals of Statistics},
	author = {Efron, Bradley and Hastie, Trevor and Johnstone, Iain and Tibshirani, Robert},
	year = {2004},
	pages = {407--499},
}

@article{st_pierre_principal-stretch-based_2023,
	title = {Principal-stretch-based constitutive neural networks autonomously discover a subclass of {Ogden} models for human brain tissue},
	volume = {4},
	issn = {26665220},
	url = {https://linkinghub.elsevier.com/retrieve/pii/S2666522023000047},
	doi = {10.1016/j.brain.2023.100066},
	language = {en},
	urldate = {2023-06-12},
	journal = {Brain Multiphysics},
	author = {St. Pierre, Skyler and Linka, Kevin and Kuhl, Ellen},
	year = {2023},
	pages = {100066},
}

@article{schroder_invariant_2003,
	title = {Invariant formulation of hyperelastic transverse isotropy based on polyconvex free energy functions},
	volume = {40},
	issn = {00207683},
	url = {https://linkinghub.elsevier.com/retrieve/pii/S0020768302004584},
	doi = {10.1016/S0020-7683(02)00458-4},
	language = {en},
	number = {2},
	urldate = {2023-01-23},
	journal = {International Journal of Solids and Structures},
	author = {Schröder, Jörg and Neff, Patrizio},
	month = jan,
	year = {2003},
	pages = {401--445},
}

@article{ogden_large_1972,
	title = {Large deformation isotropic elasticity – on the correlation of theory and experiment for incompressible rubberlike solids},
	language = {en},
	number = {326},
	journal = {Proceedings of the Royal Society of London. Series A, Mathematical and Physical Sciences,},
	author = {Ogden, Raymond William},
	year = {1972},
	pages = {565--584},
}

@misc{anton_physics-informed_2022,
	title = {Physics-{Informed} {Neural} {Networks} for {Material} {Model} {Calibration} from {Full}-{Field} {Displacement} {Data}},
	url = {http://arxiv.org/abs/2212.07723},
	language = {en},
	urldate = {2022-12-19},
	publisher = {arXiv},
	author = {Anton, David and Wessels, Henning},
	month = dec,
	year = {2022},
	note = {arXiv:2212.07723 [cs]},
}

@article{thakolkaran_nn-euclid_2022,
	title = {{NN}-{EUCLID}: {Deep}-learning hyperelasticity without stress data},
	volume = {169},
	issn = {00225096},
	shorttitle = {{NN}-{EUCLID}},
	url = {https://linkinghub.elsevier.com/retrieve/pii/S0022509622002538},
	doi = {10.1016/j.jmps.2022.105076},
	language = {en},
	urldate = {2022-10-01},
	journal = {Journal of the Mechanics and Physics of Solids},
	author = {Thakolkaran, Prakash and Joshi, Akshay and Zheng, Yiwen and Flaschel, Moritz and De Lorenzis, Laura and Kumar, Siddhant},
	month = dec,
	year = {2022},
	pages = {105076},
}

@article{flaschel_unsupervised_2021,
	title = {Unsupervised discovery of interpretable hyperelastic constitutive laws},
	volume = {381},
	doi = {10.1016/j.cma.2021.113852},
	language = {en},
	journal = {Computer Methods in Applied Mechanics and Engineering},
	author = {Flaschel, Moritz and Kumar, Siddhant and De Lorenzis, Laura},
	month = aug,
	year = {2021},
	pages = {113852},
}

@article{wang_inference_2021,
	title = {Inference of deformation mechanisms and constitutive response of soft material surrogates of biological tissue by full-field characterization and data-driven variational system identification},
	volume = {153},
	issn = {00225096},
	url = {https://linkinghub.elsevier.com/retrieve/pii/S0022509621001459},
	doi = {10.1016/j.jmps.2021.104474},
	language = {en},
	urldate = {2021-06-23},
	journal = {Journal of the Mechanics and Physics of Solids},
	author = {Wang, Z. and Estrada, J.B. and Arruda, E.M. and Garikipati, K.},
	month = aug,
	year = {2021},
	pages = {104474},
}

@article{kirchdoerfer_data-driven_2016,
	title = {Data-driven computational mechanics},
	volume = {304},
	issn = {00457825},
	url = {https://linkinghub.elsevier.com/retrieve/pii/S0045782516300238},
	doi = {10.1016/j.cma.2016.02.001},
	language = {en},
	urldate = {2020-09-09},
	journal = {Computer Methods in Applied Mechanics and Engineering},
	author = {Kirchdoerfer, T. and Ortiz, M.},
	month = jun,
	year = {2016},
	pages = {81--101},
}

@article{ghaboussi_knowledgebased_1991,
	title = {Knowledge‐{Based} {Modeling} of {Material} {Behavior} with {Neural} {Networks}},
	volume = {117},
	issn = {0733-9399, 1943-7889},
	url = {http://ascelibrary.org/doi/10.1061/%28ASCE%290733-9399%281991%29117%3A1%28132%29},
	doi = {10.1061/(ASCE)0733-9399(1991)117:1(132)},
	language = {en},
	number = {1},
	urldate = {2020-08-31},
	journal = {Journal of Engineering Mechanics},
	author = {Ghaboussi, J. and Garrett, J. H. and Wu, X.},
	month = jan,
	year = {1991},
	pages = {132--153},
}

@article{treloar_stress-strain_1944,
	title = {Stress-strain data for vulcanised rubber under various types of deformation},
	volume = {40},
	issn = {0014-7672},
	url = {http://xlink.rsc.org/?DOI=tf9444000059},
	doi = {10.1039/tf9444000059},
	language = {en},
	urldate = {2021-11-25},
	journal = {Transactions of the Faraday Society},
	author = {Treloar, L. R. G.},
	year = {1944},
	pages = {59},
}

@article{roux_optimal_2020,
	title = {Optimal procedure for the identification of constitutive parameters from experimentally measured displacement fields},
	volume = {184},
	issn = {00207683},
	url = {https://linkinghub.elsevier.com/retrieve/pii/S0020768318304542},
	doi = {10.1016/j.ijsolstr.2018.11.008},
	language = {en},
	urldate = {2021-11-16},
	journal = {International Journal of Solids and Structures},
	author = {Roux, Stéphane and Hild, François},
	month = feb,
	year = {2020},
	pages = {14--23},
}

@article{hartmann_polyconvexity_2003,
	title = {Polyconvexity of generalized polynomial-type hyperelastic strain energy functions for near-incompressibility},
	volume = {40},
	issn = {00207683},
	url = {https://linkinghub.elsevier.com/retrieve/pii/S0020768303000866},
	doi = {10.1016/S0020-7683(03)00086-6},
	language = {en},
	number = {11},
	urldate = {2021-11-05},
	journal = {International Journal of Solids and Structures},
	author = {Hartmann, Stefan and Neff, Patrizio},
	month = jun,
	year = {2003},
	pages = {2767--2791},
}

@article{grediac_principle_1989,
	title = {Principle of virtual work and identification},
	number = {309(1)},
	journal = {Comptes Rendus de L Academie des Sciences Serie Ii},
	author = {Grédiac, M.},
	year = {1989},
	pages = {1--5},
}

@article{ibanez_data-driven_2017,
	title = {Data-driven non-linear elasticity: constitutive manifold construction and problem discretization},
	volume = {60},
	issn = {0178-7675, 1432-0924},
	shorttitle = {Data-driven non-linear elasticity},
	url = {http://link.springer.com/10.1007/s00466-017-1440-1},
	doi = {10.1007/s00466-017-1440-1},
	language = {en},
	number = {5},
	urldate = {2020-08-24},
	journal = {Computational Mechanics},
	author = {Ibañez, Ruben and Borzacchiello, Domenico and Aguado, Jose Vicente and Abisset-Chavanne, Emmanuelle and Cueto, Elias and Ladeveze, Pierre and Chinesta, Francisco},
	month = nov,
	year = {2017},
	pages = {813--826},
}

@article{hild_digital_2006,
	title = {Digital {Image} {Correlation}: from {Displacement} {Measurement} to {Identification} of {Elastic} {Properties} - a {Review}},
	volume = {42},
	issn = {00392103, 14751305},
	shorttitle = {Digital {Image} {Correlation}},
	url = {http://doi.wiley.com/10.1111/j.1475-1305.2006.00258.x},
	doi = {10.1111/j.1475-1305.2006.00258.x},
	language = {en},
	number = {2},
	urldate = {2020-07-16},
	journal = {Strain},
	author = {Hild, F. and Roux, S.},
	month = may,
	year = {2006},
	pages = {69--80},
}

@article{avril_overview_2008,
	title = {Overview of {Identification} {Methods} of {Mechanical} {Parameters} {Based} on {Full}-field {Measurements}},
	volume = {48},
	issn = {0014-4851, 1741-2765},
	url = {http://link.springer.com/10.1007/s11340-008-9148-y},
	doi = {10.1007/s11340-008-9148-y},
	language = {en},
	number = {4},
	urldate = {2020-07-16},
	journal = {Experimental Mechanics},
	author = {Avril, Stéphane and Bonnet, Marc and Bretelle, Anne-Sophie and Grédiac, Michel and Hild, François and Ienny, Patrick and Latourte, Félix and Lemosse, Didier and Pagano, Stéphane and Pagnacco, Emmanuel and Pierron, Fabrice},
	month = aug,
	year = {2008},
	pages = {381--402},
}

@inproceedings{asad_mechanics-informed_2022,
	address = {San Diego, CA \& Virtual},
	title = {A {Mechanics}-{Informed} {Artificial} {Neural} {Network} {Approach} in {Data}-{Driven} {Constitutive} {Modeling}},
	isbn = {978-1-62410-631-6},
	url = {https://arc.aiaa.org/doi/10.2514/6.2022-0100},
	doi = {10.2514/6.2022-0100},
	language = {en},
	urldate = {2022-05-19},
	booktitle = {{AIAA} {SCITECH} 2022 {Forum}},
	publisher = {American Institute of Aeronautics and Astronautics},
	author = {As'ad, Faisal and Avery, Philip and Farhat, Charbel},
	month = jan,
	year = {2022},
}

@article{klein_polyconvex_2022,
	title = {Polyconvex anisotropic hyperelasticity with neural networks},
	volume = {159},
	issn = {00225096},
	url = {https://linkinghub.elsevier.com/retrieve/pii/S0022509621003215},
	doi = {10.1016/j.jmps.2021.104703},
	language = {en},
	urldate = {2022-05-16},
	journal = {Journal of the Mechanics and Physics of Solids},
	author = {Klein, Dominik K. and Fernández, Mauricio and Martin, Robert J. and Neff, Patrizio and Weeger, Oliver},
	month = feb,
	year = {2022},
	pages = {104703},
}

@article{ball_convexity_1976,
	title = {Convexity conditions and existence theorems in nonlinear elasticity},
	volume = {63},
	issn = {0003-9527, 1432-0673},
	url = {http://link.springer.com/10.1007/BF00279992},
	doi = {10.1007/BF00279992},
	language = {en},
	number = {4},
	urldate = {2022-03-10},
	journal = {Archive for Rational Mechanics and Analysis},
	author = {Ball, John M.},
	month = dec,
	year = {1976},
	pages = {337--403},
}

@article{linka_new_2023,
	title = {A new family of {Constitutive} {Artificial} {Neural} {Networks} towards automated model discovery},
	volume = {403},
	issn = {00457825},
	url = {https://linkinghub.elsevier.com/retrieve/pii/S0045782522006867},
	doi = {10.1016/j.cma.2022.115731},
	language = {en},
	urldate = {2024-11-13},
	journal = {Computer Methods in Applied Mechanics and Engineering},
	author = {Linka, Kevin and Kuhl, Ellen},
	month = jan,
	year = {2023},
	pages = {115731},
}

@article{linden_neural_2023,
	title = {Neural networks meet hyperelasticity: {A} guide to enforcing physics},
	volume = {179},
	issn = {00225096},
	shorttitle = {Neural networks meet hyperelasticity},
	url = {https://linkinghub.elsevier.com/retrieve/pii/S0022509623001679},
	doi = {10.1016/j.jmps.2023.105363},
	language = {en},
	urldate = {2024-11-13},
	journal = {Journal of the Mechanics and Physics of Solids},
	author = {Linden, Lennart and Klein, Dominik K. and Kalina, Karl A. and Brummund, Jörg and Weeger, Oliver and Kästner, Markus},
	month = oct,
	year = {2023},
	pages = {105363},
}

@article{fuhg_review_2024,
	title = {A {Review} on {Data}-{Driven} {Constitutive} {Laws} for {Solids}},
	issn = {1134-3060, 1886-1784},
	url = {https://link.springer.com/10.1007/s11831-024-10196-2},
	doi = {10.1007/s11831-024-10196-2},
	language = {en},
	urldate = {2024-11-13},
	journal = {Archives of Computational Methods in Engineering},
	author = {Fuhg, Jan N. and Anantha Padmanabha, Govinda and Bouklas, Nikolaos and Bahmani, Bahador and Sun, WaiChing and Vlassis, Nikolaos N. and Flaschel, Moritz and Carrara, Pietro and De Lorenzis, Laura},
	month = nov,
	year = {2024},
}

@article{rosenkranz_comparative_2023,
	title = {A comparative study on different neural network architectures to model inelasticity},
	volume = {124},
	issn = {0029-5981, 1097-0207},
	url = {https://onlinelibrary.wiley.com/doi/10.1002/nme.7319},
	doi = {10.1002/nme.7319},
	language = {en},
	number = {21},
	urldate = {2024-11-08},
	journal = {International Journal for Numerical Methods in Engineering},
	author = {Rosenkranz, Max and Kalina, Karl A. and Brummund, Jörg and Kästner, Markus},
	month = nov,
	year = {2023},
	pages = {4802--4840},
}

@article{masi_evolution_2023,
	title = {Evolution {TANN} and the identification of internal variables and evolution equations in solid mechanics},
	volume = {174},
	issn = {00225096},
	url = {https://linkinghub.elsevier.com/retrieve/pii/S0022509623000492},
	doi = {10.1016/j.jmps.2023.105245},
	language = {en},
	urldate = {2024-10-04},
	journal = {Journal of the Mechanics and Physics of Solids},
	author = {Masi, Filippo and Stefanou, Ioannis},
	month = may,
	year = {2023},
	pages = {105245},
}

@article{benady_unsupervised_2024,
	title = {Unsupervised learning of history-dependent constitutive material laws with thermodynamically-consistent neural networks in the modified {Constitutive} {Relation} {Error} framework},
	volume = {425},
	issn = {00457825},
	url = {https://linkinghub.elsevier.com/retrieve/pii/S0045782524002238},
	doi = {10.1016/j.cma.2024.116967},
	language = {en},
	urldate = {2024-09-26},
	journal = {Computer Methods in Applied Mechanics and Engineering},
	author = {Benady, Antoine and Baranger, Emmanuel and Chamoin, Ludovic},
	month = may,
	year = {2024},
	pages = {116967},
}

@misc{mcculloch_automated_2024,
	title = {Automated model discovery for textile structures: {The} unique mechanical signature of warp knitted fabrics},
	copyright = {http://creativecommons.org/licenses/by/4.0/},
	shorttitle = {Automated model discovery for textile structures},
	url = {http://biorxiv.org/lookup/doi/10.1101/2024.07.26.605392},
	doi = {10.1101/2024.07.26.605392},
	language = {en},
	urldate = {2024-09-18},
	author = {McCulloch, Jeremy A. and Kuhl, Ellen},
	month = jul,
	year = {2024},
}

@misc{jadoon_automated_2024,
	title = {Automated model discovery of finite strain elastoplasticity from uniaxial experiments},
	url = {http://arxiv.org/abs/2408.14615},
	language = {en},
	urldate = {2024-09-17},
	publisher = {arXiv},
	author = {Jadoon, Asghar A. and Meyer, Knut A. and Fuhg, Jan N.},
	month = aug,
	year = {2024},
	note = {arXiv:2408.14615 [cs]},
}

@article{wiesheier_versatile_2024,
	title = {Versatile data-adaptive hyperelastic energy functions for soft materials},
	volume = {430},
	issn = {00457825},
	url = {https://linkinghub.elsevier.com/retrieve/pii/S004578252400464X},
	doi = {10.1016/j.cma.2024.117208},
	language = {en},
	urldate = {2024-09-05},
	journal = {Computer Methods in Applied Mechanics and Engineering},
	author = {Wiesheier, Simon and Moreno-Mateos, Miguel Angel and Steinmann, Paul},
	month = oct,
	year = {2024},
	pages = {117208},
}

@article{holthusen_theory_2024,
	title = {Theory and implementation of inelastic {Constitutive} {Artificial} {Neural} {Networks}},
	volume = {428},
	issn = {00457825},
	url = {https://linkinghub.elsevier.com/retrieve/pii/S0045782524003190},
	doi = {10.1016/j.cma.2024.117063},
	language = {en},
	urldate = {2024-09-02},
	journal = {Computer Methods in Applied Mechanics and Engineering},
	author = {Holthusen, Hagen and Lamm, Lukas and Brepols, Tim and Reese, Stefanie and Kuhl, Ellen},
	month = aug,
	year = {2024},
	pages = {117063},
}

@book{hastie_elements_2009,
	address = {New York, NY},
	series = {Springer {Series} in {Statistics}},
	title = {The {Elements} of {Statistical} {Learning}: {Data} {Mining}, {Inference}, and {Prediction}},
	isbn = {978-0-387-84857-0 978-0-387-84858-7},
	url = {http://link.springer.com/10.1007/978-0-387-84858-7},
	language = {en},
	urldate = {2022-12-20},
	publisher = {Springer New York},
	author = {Hastie, Trevor and Tibshirani, Robert and Friedman, Jerome},
	year = {2009},
	doi = {10.1007/978-0-387-84858-7},
}

@phdthesis{flaschel_automated_2023-2,
	title = {Automated {Discovery} of {Material} {Models} in {Continuum} {Solid} {Mechanics}},
	url = {http://hdl.handle.net/20.500.11850/602750},
	language = {en},
	urldate = {2023-03-14},
	school = {ETH Zurich},
	author = {Flaschel, Moritz},
	collaborator = {De Lorenzis, Laura and Kumar, Siddhant and Steinmann, Paul},
	year = {2023},
	doi = {10.3929/ETHZ-B-000602750},
}

@article{abdusalamov_automatic_2023,
	title = {Automatic generation of interpretable hyperelastic material models by symbolic regression},
	issn = {0029-5981, 1097-0207},
	url = {https://onlinelibrary.wiley.com/doi/10.1002/nme.7203},
	doi = {10.1002/nme.7203},
	language = {en},
	urldate = {2023-01-30},
	journal = {International Journal for Numerical Methods in Engineering},
	author = {Abdusalamov, Rasul and Hillgärtner, Markus and Itskov, Mikhail},
	month = jan,
	year = {2023},
	pages = {nme.7203},
}

@article{boehler_irreducible_1977,
	title = {On {Irreducible} {Representations} for {Isotropic} {Scalar} {Functions}},
	volume = {57},
	copyright = {http://onlinelibrary.wiley.com/termsAndConditions\#vor},
	issn = {0044-2267, 1521-4001},
	url = {https://onlinelibrary.wiley.com/doi/10.1002/zamm.19770570608},
	doi = {10.1002/zamm.19770570608},
	language = {en},
	number = {6},
	urldate = {2025-03-06},
	journal = {ZAMM - Journal of Applied Mathematics and Mechanics / Zeitschrift für Angewandte Mathematik und Mechanik},
	author = {Boehler, J. P.},
	month = jan,
	year = {1977},
	pages = {323--327},
}

@article{flaschel_convex_2025,
	title = {Convex neural networks learn generalized standard material models},
	volume = {200},
	issn = {00225096},
	url = {https://linkinghub.elsevier.com/retrieve/pii/S0022509625000791},
	doi = {10.1016/j.jmps.2025.106103},
	language = {en},
	urldate = {2025-03-13},
	journal = {Journal of the Mechanics and Physics of Solids},
	author = {Flaschel, Moritz and Steinmann, Paul and De Lorenzis, Laura and Kuhl, Ellen},
	month = jul,
	year = {2025},
	pages = {106103},
}

@misc{geuken_input_2025,
	title = {Input convex neural networks: universal approximation theorem and implementation for isotropic polyconvex hyperelastic energies},
	shorttitle = {Input convex neural networks},
	url = {http://arxiv.org/abs/2502.08534},
	doi = {10.48550/arXiv.2502.08534},
	language = {en},
	urldate = {2025-03-26},
	publisher = {arXiv},
	author = {Geuken, Gian-Luca and Kurzeja, Patrick and Wiedemann, David and Mosler, Jörn},
	month = feb,
	year = {2025},
	note = {arXiv:2502.08534 [cs]},
}

@misc{flaschel_non-smooth_2025,
	title = {Non-smooth optimization meets automated material model discovery},
	url = {http://arxiv.org/abs/2507.10196},
	doi = {10.48550/arXiv.2507.10196},
	language = {en},
	urldate = {2025-07-16},
	publisher = {arXiv},
	author = {Flaschel, Moritz and Hastie, Trevor and Kuhl, Ellen},
	month = jul,
	year = {2025},
	note = {arXiv:2507.10196 [cs]},
}

@incollection{cowin_new_2004,
	address = {Dordrecht},
	title = {A new {Constitutive} {Framework} for {Arterial} {Wall} {Mechanics} and a {Comparative} {Study} of {Material} {Models}},
	isbn = {978-1-4020-0220-5},
	url = {http://link.springer.com/10.1007/0-306-48389-0_1},
	language = {en},
	urldate = {2025-08-06},
	booktitle = {Cardiovascular {Soft} {Tissue} {Mechanics}},
	publisher = {Kluwer Academic Publishers},
	author = {Holzapfel, Gerhard A. and Gasser, Thomas C. and Ogden, Ray W.},
	editor = {Cowin, Stephen C. and Humphrey, Jay D.},
	year = {2004},
	doi = {10.1007/0-306-48389-0_1},
	pages = {1--48},
}

@article{martonova_automated_2024,
	title = {Automated model discovery for human cardiac tissue: {Discovering} the best model and parameters},
	volume = {428},
	issn = {00457825},
	shorttitle = {Automated model discovery for human cardiac tissue},
	url = {https://linkinghub.elsevier.com/retrieve/pii/S0045782524003347},
	doi = {10.1016/j.cma.2024.117078},
	language = {en},
	urldate = {2025-08-08},
	journal = {Computer Methods in Applied Mechanics and Engineering},
	author = {Martonová, Denisa and Peirlinck, Mathias and Linka, Kevin and Holzapfel, Gerhard A. and Leyendecker, Sigrid and Kuhl, Ellen},
	month = aug,
	year = {2024},
	pages = {117078},
}

@misc{martonova_generalized_2025,
	title = {Generalized invariants meet constitutive neural networks: {A} novel framework for hyperelastic materials},
	shorttitle = {Generalized invariants meet constitutive neural networks},
	url = {http://arxiv.org/abs/2508.12063},
	doi = {10.48550/arXiv.2508.12063},
	language = {en},
	urldate = {2025-08-22},
	publisher = {arXiv},
	author = {Martonová, Denisa and Goriely, Alain and Kuhl, Ellen},
	month = aug,
	year = {2025},
	note = {arXiv:2508.12063 [cond-mat]},
}

@article{kalina_neural_2025,
	title = {Neural networks meet anisotropic hyperelasticity: {A} framework based on generalized structure tensors and isotropic tensor functions},
	volume = {437},
	issn = {00457825},
	shorttitle = {Neural networks meet anisotropic hyperelasticity},
	url = {https://linkinghub.elsevier.com/retrieve/pii/S0045782524009812},
	doi = {10.1016/j.cma.2024.117725},
	language = {en},
	urldate = {2025-09-29},
	journal = {Computer Methods in Applied Mechanics and Engineering},
	author = {Kalina, Karl A. and Brummund, Jörg and Sun, WaiChing and Kästner, Markus},
	month = mar,
	year = {2025},
	pages = {117725},
}

@article{romer_reduced_2025,
	title = {Reduced and {All}-{At}-{Once} {Approaches} for {Model} {Calibration} and {Discovery} in {Computational} {Solid} {Mechanics}},
	volume = {77},
	issn = {0003-6900, 2379-0407},
	url = {https://asmedigitalcollection.asme.org/appliedmechanicsreviews/article/77/4/040801/1201974/Reduced-and-All-At-Once-Approaches-for-Model},
	doi = {10.1115/1.4066118},
	language = {en},
	number = {4},
	urldate = {2025-10-29},
	journal = {Applied Mechanics Reviews},
	author = {Römer, Ulrich and Hartmann, Stefan and Tröger, Jendrik-Alexander and Anton, David and Wessels, Henning and Flaschel, Moritz and De Lorenzis, Laura},
	month = jul,
	year = {2025},
	pages = {040801},
}

@article{flaschel_material_2026,
	title = {Material {Fingerprinting}: {A} shortcut to material model discovery without solving optimization problems},
	volume = {450},
	issn = {00457825},
	shorttitle = {Material {Fingerprinting}},
	url = {https://linkinghub.elsevier.com/retrieve/pii/S004578252500845X},
	doi = {10.1016/j.cma.2025.118573},
	language = {en},
	urldate = {2025-12-02},
	journal = {Computer Methods in Applied Mechanics and Engineering},
	author = {Flaschel, Moritz and Martonová, Denisa and Veil, Carina and Kuhl, Ellen},
	month = mar,
	year = {2026},
	pages = {118573},
}

@article{martonova_material_2026,
	title = {Material {Fingerprinting} for rapid discovery of hyperelastic models: {First} experimental validation},
	volume = {208},
	issn = {00225096},
	shorttitle = {Material {Fingerprinting} for rapid discovery of hyperelastic models},
	url = {https://linkinghub.elsevier.com/retrieve/pii/S0022509625004375},
	doi = {10.1016/j.jmps.2025.106463},
	language = {en},
	urldate = {2025-12-12},
	journal = {Journal of the Mechanics and Physics of Solids},
	author = {Martonová, Denisa and Kuhl, Ellen and Flaschel, Moritz},
	month = feb,
	year = {2026},
	pages = {106463},
}

@phdthesis{klein_polyconvex_2025,
	title = {Polyconvex {Hyperelasticity} with {Neural} {Networks}: {On} {Invariant}- and {Coordinate}-{Based} {Models}, {Benefits} and {Limitations}},
	copyright = {Creative Commons Attribution Share Alike 4.0 International},
	shorttitle = {Polyconvex {Hyperelasticity} with {Neural} {Networks}},
	url = {https://tuprints.ulb.tu-darmstadt.de/handle/tuda/14802},
	language = {de},
	urldate = {2026-01-06},
	school = {Universitäts- und Landesbibliothek Darmstadt},
	author = {Klein, Dominik Klemens},
	collaborator = {{Universitäts- und Landesbibliothek Darmstadt}},
	year = {2025},
	doi = {10.26083/TUDA-7577},
}

@article{urreaquintero_automated_2026,
	title = {Automated constitutive model discovery by pairing sparse regression algorithms with model selection criteria},
	volume = {449},
	issn = {00457825},
	url = {https://linkinghub.elsevier.com/retrieve/pii/S0045782525008230},
	doi = {10.1016/j.cma.2025.118551},
	language = {en},
	urldate = {2026-01-07},
	journal = {Computer Methods in Applied Mechanics and Engineering},
	author = {Urrea–Quintero, Jorge–Humberto and Anton, David and De Lorenzis, Laura and Wessels, Henning},
	month = feb,
	year = {2026},
	pages = {118551},
}

@misc{flaschel_unsupervised_2026,
	title = {Unsupervised {Material} {Fingerprinting}: {Ultra}-fast hyperelastic model discovery from full-field experimental measurements},
	shorttitle = {Unsupervised {Material} {Fingerprinting}},
	url = {http://arxiv.org/abs/2601.14965},
	doi = {10.48550/arXiv.2601.14965},
	language = {en},
	urldate = {2026-01-22},
	publisher = {arXiv},
	author = {Flaschel, Moritz and Moreno-Mateos, Miguel Angel and Wiesheier, Simon and Steinmann, Paul and Kuhl, Ellen},
	month = jan,
	year = {2026},
	note = {arXiv:2601.14965 [cs]},
}

@article{linka_automated_2023-1,
	title = {Automated model discovery for skin: {Discovering} the best model, data, and experiment},
	volume = {410},
	issn = {00457825},
	shorttitle = {Automated model discovery for skin},
	url = {https://linkinghub.elsevier.com/retrieve/pii/S0045782523001317},
	doi = {10.1016/j.cma.2023.116007},
	language = {en},
	urldate = {2026-02-03},
	journal = {Computer Methods in Applied Mechanics and Engineering},
	author = {Linka, Kevin and Buganza Tepole, Adrian and Holzapfel, Gerhard A. and Kuhl, Ellen},
	month = may,
	year = {2023},
	pages = {116007},
}

@book{ciarlet_mathematical_1988,
	series = {Studies in {Mathematics} and its {Applications}},
	title = {Mathematical {Elasticity}, {Volume} {I}: {Three}-{Dimensional} {Elasticity}},
	volume = {20},
	isbn = {978-0-444-70259-4},
	publisher = {North-Holland},
	author = {Ciarlet, Philippe G.},
	year = {1988},
}

@book{horn_matrix_2012,
	address = {New York},
	title = {Matrix analysis},
	publisher = {Cambridge University Press},
	author = {Horn, Roger A. and Johnson, Charles R.},
	year = {2012},
}

@article{boehler_simple_1979,
	title = {A {Simple} {Derivation} of {Representations} for {Non}‐{Polynomial} {Constitutive} {Equations} in {Some} {Cases} of {Anisotropy}},
	volume = {59},
	copyright = {http://onlinelibrary.wiley.com/termsAndConditions\#vor},
	issn = {0044-2267, 1521-4001},
	url = {https://onlinelibrary.wiley.com/doi/10.1002/zamm.19790590403},
	doi = {10.1002/zamm.19790590403},
	language = {en},
	number = {4},
	urldate = {2026-02-09},
	journal = {ZAMM - Journal of Applied Mathematics and Mechanics / Zeitschrift für Angewandte Mathematik und Mechanik},
	author = {Boehler, Jean‐Paul},
	month = jan,
	year = {1979},
	pages = {157--167},
}

@book{gdoutos_mechanical_2024,
	address = {Cham},
	series = {Solid {Mechanics} and {Its} {Applications}},
	title = {Mechanical {Testing} of {Materials}},
	volume = {275},
	copyright = {https://www.springernature.com/gp/researchers/text-and-data-mining},
	isbn = {978-3-031-45989-4 978-3-031-45990-0},
	url = {https://link.springer.com/10.1007/978-3-031-45990-0},
	language = {en},
	urldate = {2026-02-09},
	publisher = {Springer Nature Switzerland},
	author = {Gdoutos, Emmanuel and Konsta-Gdoutos, Maria},
	year = {2024},
	doi = {10.1007/978-3-031-45990-0},
}

@article{tepole_polyconvex_2025,
	title = {Polyconvex physics-augmented neural network constitutive models in principal stretches},
	volume = {320},
	issn = {00207683},
	url = {https://linkinghub.elsevier.com/retrieve/pii/S0020768325002550},
	doi = {10.1016/j.ijsolstr.2025.113469},
	language = {en},
	urldate = {2026-04-21},
	journal = {International Journal of Solids and Structures},
	author = {Tepole, Adrian Buganza and Jadoon, Asghar Arshad and Rausch, Manuel and Fuhg, Jan Niklas},
	month = sep,
	year = {2025},
	pages = {113469},
}

@article{vijayakumaran_consistent_2025,
	title = {Consistent machine learning for topology optimization with microstructure-dependent neural network material models},
	volume = {196},
	issn = {00225096},
	url = {https://linkinghub.elsevier.com/retrieve/pii/S0022509624004812},
	doi = {10.1016/j.jmps.2024.106015},
	language = {en},
	urldate = {2026-05-08},
	journal = {Journal of the Mechanics and Physics of Solids},
	author = {Vijayakumaran, Harikrishnan and Russ, Jonathan B. and Paulino, Glaucio H. and Bessa, Miguel A.},
	month = mar,
	year = {2025},
	pages = {106015},
}
% \bibliography{bib_Moritz,bib_Hagen,bib_Denisa}

%%%%%%%%%%%%%%%%%%%%%%%%%%%%%%%%%%%%%%%%%%%%%%%%%%%%%%%%%%%%%%%%%%%%%%%%%%%%%%%%%%%%%%%%%%%%%%%%%%%%%%%%%%%%%%%%%%%%%%%

\end{document}